\documentclass[fleqn,usenatbib]{mnras}

\usepackage{newtxtext,newtxmath}

\usepackage[T1]{fontenc}

\DeclareRobustCommand{\VAN}[3]{#2}
\let\VANthebibliography\thebibliography
\def\thebibliography{\DeclareRobustCommand{\VAN}[3]{##3}\VANthebibliography}

\usepackage{graphicx}	
\usepackage{amsmath}	

\title[The first DLA silhouette image]{The first direct imaging of the silhouette of a damped Lyman $\alpha$ system along the line-of-sight to a background galaxy}

\author[F. Komori et al.]{
Fuga Komori,$^{1}$
Akio K. Inoue,$^{1,2}$\thanks{E-mail: akinoue@aoni.waseda.jp (AKI)}
Ken Mawatari,$^{1,2}$
Yuma Sugahara,$^{1,2}$
Hideki Umehata,$^{3,4}$
Rhythm \newauthor Shimakawa,$^{5}$
Satoshi Yamanaka,$^{6}$
Takuya Hashimoto,$^{7,8}$
Jorryt Matthee,$^{9}$
and
Toru Misawa$^{10}$
\\
$^{1}$Department of Pure and Applied Physics, Graduate School of Advanced Science and Engineering, Faculty of Science and Engineering, Waseda University,\\3-4-1, Okubo, Shinjuku, Tokyo, 169-8555, Japan\\
$^{2}$Waseda Research Institute for Science and Engineering, Faculty of Science and Engineering, Waseda University, 3-4-1, Okubo, Shinjuku, Tokyo, 169-8555, Japan\\
$^{3}$Institute for Advanced Research, Nagoya University, Furocho, Chikusa, Nagoya 464-8602, Japan\\
$^{4}$Department of Physics, Graduate School of Science, Nagoya University, Furocho, Chikusa, Nagoya 464-8602, Japan\\
$^{5}$Waseda Institute for Advanced Study (WIAS), Waseda University, 1-21-1, Nishi-Waseda, Shinjuku, Tokyo 169-0051, Japan\\
$^{6}$General Education Department, National Institute of Technology, Toba College, 1-1 Ikegami-cho, Toba, Mie 517-8501, Japan\\
$^{7}$Graduate School of Pure and Applied Sciences, University of Tsukuba, 1-1-1 Tennodai, Tsukuba, Ibaraki 305-8571, Japan\\
$^{8}$Tomonaga Center for the History of the Universe, University of Tsukuba, 1-1-1 Tennodai, Tsukuba, Ibaraki 305-8571, Japan\\
$^{9}$Institute of Science and Technology Austria (ISTA), Am Campus 1, 3400 Klosterneuburg, Austria\\
$^{10}$Center for General Education, Shinshu University, 3-1-1 Asahi, Matsumoto, Nagano 390-8621, Japan\\
}

\date{Accepted 2025 September 22. Received 2025 September 22; in original form 2025 May 17.}

\pubyear{\the\year{}}

\begin{document}
\label{firstpage}
\pagerange{\pageref{firstpage}--\pageref{lastpage}}
\maketitle

\begin{abstract}
The H~{\sc i} gas distribution in damped Lyman $\alpha$ absorbers (DLAs) has remained elusive due to the point-source nature of background quasar emission.
Observing DLAs against spatially extended background galaxies provides a new method for constraining their size and structure.
Using the Keck Cosmic Web Imager, we present the first ``silhouette'' image of a DLA at $z=3.34$, identified in the spectrum of a background galaxy at $z=3.61$. 
Although the silhouette remains unresolved due to limited spatial resolution, this represents a successful proof-of-concept for studying DLA morphology using extended background sources.
Possible residual emission in the DLA trough suggests an optical depth contrast exceeding $10^7$ in the internal structure, implying a sharp edge or patchy structure.
A Lyman $\alpha$ emitter (LAE) at $z_{\rm LAE}=3.3433\pm0.0005$, consistent with the DLA redshift, is detected at an angular separation of $1.''73\pm0.''28$ ($12.9\pm2.1$ kpc).
The DLA is surrounded by three galaxies within 140 kpc in projected distance and 500 km s$^{-1}$ in line-of-sight velocity, indicating that it resides in the circumgalactic medium of the LAE or within a galaxy group/protocluster environment.
An O~{\sc i} $\lambda1302$ absorption at $z_{\rm OI}=3.3288\pm0.0004$ is also detected along the line of sight.
This absorber may trace metal-enriched outflow from the LAE or a gas-rich galaxy exhibiting the highest star formation activity among the surrounding galaxies.
Future large spectroscopic surveys of galaxies will expand such a DLA sample, and three-dimensional spectroscopy for it will shed new light on the role of intergalactic dense gas in galaxy formation and evolution.
\end{abstract}

\begin{keywords}
 (galaxies:) intergalactic medium -- (galaxies:) quasars: absorption lines -- galaxies: evolution
\end{keywords}



\section{Introduction}\label{sec:intro}

The supply of neutral hydrogen (H~{\sc i}) gas from the circumgalactic medium (CGM) and the intergalactic medium (IGM) is essential for galaxies to continue star formation \citep[e.g.,][]{2005ApJ...635..123P,2017ARA&A..55..389T}. 
Therefore, studying the relationship between the gas surrounding galaxies and the stellar component in galaxies is crucial for understanding galaxy formation and evolution. 
The spectra of extremely bright extragalactic sources, known as quasars, confirm the presence of high-density H~{\sc i} gas clouds in the intervening medium along the line of sight, known as damped Lyman $\alpha$ systems (DLAs), defined by an H~{\sc i} column density of $N_{\rm HI} > 2\times10^{20}$ cm$^{-2}$ \citep[e.g.,][]{2005ARA&A..43..861W,2009RvMP...81.1405M}.
Since the H~{\sc i} column density of DLAs is similar to that of galactic disks, DLAs are thought to be the outer part of galactic disks and/or the material in the CGM \citep[e.g.,][]{2005ARA&A..43..861W}. 
DLAs at $z\sim3$ contain the H~{\sc i} gas mass of 20-50\% of the stellar mass in the present universe \citep[e.g.,][]{1995ApJ...440..435L,2000ApJ...543..552S}.
Therefore, DLAs have been the focus of research on the CGM and IGM at high redshifts for many years \citep[e.g.,][]{2003MNRAS.346.1103P,2009ApJ...696.1543P,2012ApJ...755...89R}.
However, because quasars are point sources with extremely small emission regions, they provide only one-dimensional information along the line of sight. 
Therefore, the size and structure of DLAs remains elusive.

In principle, if there is more than one line of sight within the extent of a DLA, we can investigate the size and structure of the absorber.
Such examples are reported with a binary quasar pair \citep{2007MNRAS.378..801E} and gravitationally-lensed multiple images of quasars \citep{2009MNRAS.397..943M,2010MNRAS.409..679C}. 
\cite{2007MNRAS.378..801E} have found a coincident pair of DLAs 110 kpc away on the two lines of sight to the binary quasar, while this example may not be a single DLA.
On the other hand, \cite{2009MNRAS.397..943M} and \cite{2010MNRAS.409..679C} have reported the size constraint on DLAs as $\sim5$--10 kpc from a quadruply lensed quasar or a doubly lensed quasar, respectively.
They have also reported more than a 10--100 fold variation in H~{\sc i} column density within the small spatial scale probed by different lensed images, i.e. different lines of sight.
\cite{2018A&A...619A.142K} have reported the detection of H$_2$ gas in the DLA along the line of sight to a doubly lensed quasar, providing the cold molecular gas size larger than the separation of the two images $>0.7$ kpc.
\cite{2015ApJ...808...38R} presented a statistical sample of 40 DLAs probed by quasar pairs, and concluded that the spatial size of DLAs rarely extends beyond scales of $>10$ kpc.
More recently, \cite{2021AJ....162..175O} have reported a significant difference in Mg~{\sc ii} absorption in the same lensed quasar as that observed by \cite{2009MNRAS.397..943M}, and \cite{2021AJ....161...90C} have reached similar conclusions with a different lensed quasar.
\cite{2022A&A...657A.113L} have reported a case of proximity DLA in another lensed quasar found with {\it Gaia}.

An alternative approach to studying the structure of DLAs is to look for any residual emission in the dark trough of the DLA spectra.
\cite{2014ApJ...793..139C} have reported the detection of the residual flux in a stack of $\sim2,000$ DLA spectra obtained from the SDSS and concluded that this is the emission of the host galaxies of the quasars, and therefore the DLA gas does not fully cover the host galaxy scale.
Partial coverage of gas in the systems associated with DLAs is also reported.
For example, \cite{2015MNRAS.448..280K} and \cite{2020MNRAS.493.5743K} report the residual flux at the bottom of the H$_2$ molecular absorption and \cite{2016MNRAS.455.2698K} report the same as C~{\sc i} absorption.
The H$_2$ molecular gas covering fraction widely distributes from $\sim10$\% to $\sim100$\% \citep{2020MNRAS.493.5743K}.
However, the H$_2$ gas is located near the broad line region of the quasar core, not in the intervening medium.
On the other hand, \cite{2009MNRAS.394L..61K} reports generally high covering fractions, $>0.4$, of H~{\sc i} gas in DLAs based on the VLBA observations at or near the redshifted 21 cm frequency.

Yet another way to constrain the size and structure of DLAs is to observe them along lines of sight to spatially extended background light sources, such as star-forming galaxies (SFGs).
Observing DLAs as foreground ``silhouettes'' in SFGs allows us to discuss the size and structure of the DLAs in a way that was impossible with quasars. 
Although there are only four cases of DLAs found in front of galaxies so far \citep{2015ApJ...812L..27C,2016ApJ...817..161M,2021ApJ...907..103D,2022Natur.606...59B}, they have provided unique constraints on the spatial extent and variation of the gas distribution in DLAs.
\cite{2015ApJ...812L..27C} report the first detection of a DLA at $z=2.391$ in the spectrum of a background galaxy at $z=2.817$ and constrain the area of the DLA gas to a few to $\sim100$ kpc$^2$.
\cite{2016ApJ...817..161M} report the second example of such a DLA at $z=3.335$ in a galaxy spectrum and constrain the area to $>1$ kpc$^2$.
\cite{2021ApJ...907..103D} report a low-$z$ example of the DLA as the foreground of a spatially resolved galaxy at $z=0.175$, placing a lower limit of the DLA gas area as $>3.3$ kpc$^2$.
\cite{2022Natur.606...59B} report a very interesting high-$z$ case, two DLAs at $z=2.056$ and $z=2.543$, in a gravitationally lensed extended galaxy at $z=2.762$.
These DLAs are widely extended to at least 238 kpc$^2$ and show spatial variations of more than an order of magnitude in the column densities of the H~{\sc i} gas and some elements such as carbon, oxygen, sulphur and silicon.

Observing DLAs on the lines of sight of galaxies has another interesting advantage over the case for quasars.
Since SFGs are more than 10--100 times fainter than quasars that outshine the host galaxies of DLAs, we may have a better chance of finding DLA host galaxies even if they are very close to the lines of sight.
Indeed, optical surveys of host galaxies of DLAs on quasar lines of sight have struggled to remove bright quasar light \citep[e.g.,][]{2017MNRAS.469.2959K}.
Recently, there have been successful observations of host galaxies and environments of DLAs, even on quasar lines of sight, using optical integral field spectroscopy \citep{2019MNRAS.487.5070M,2022MNRAS.514.6074N,2024ApJ...962...72O}, because quasar light is almost invisible in the dark trough of DLAs and Ly$\alpha$ emission at the same redshift as the DLAs emerge.
This merit of the integral field spectroscopy holds for galaxy lines of sight.

In this paper, we present observational results of integral field spectroscopy with the Keck Cosmic Web Imager (KCWI; \citealt{2018ApJ...864...93M}) on the 10-m Keck telescope for a DLA at $z=3.34$ in the spectrum of a bright background Lyman Break Galaxy (LBG) at $z=3.61$ identified by \cite{2016ApJ...817..161M}. 
Thanks to very sensitive observations under an on-source time of 24,000 seconds, we present for the first time a ``silhouette'' image of the DLA that constrains the size and structure of the H~{\sc i} gas distribution.

The rest of this paper is structured as follows:
In section 2, we describe the observations and data reduction, followed by the detailed discussion of the spectral analysis in section 3, where we report detections of a Ly$\alpha$ emitter and an O~{\sc i} absorber associated with the DLA.
In section 4, we present the DLA silhouette image and its radial profile.
In the final section 5, we discuss the host galaxy of the DLA, the nature of the DLA and the O~{\sc i} absorber, and future prospects.
This paper assumes a $\Lambda$ Cold Dark Matter cosmology with the Hubble constant of $H_0 = 70$ km s$^{-1}$ Mpc$^{-1}$, the matter density parameter of $\Omega_m = 0.3$, and the dark energy density parameter of $\Omega_\Lambda = 0.7$. 
The baryon density parameter is assumed to be $\Omega_b h^2 = 0.0224$ \citep{2020A&A...641A...6P}, where $h=H_0/100$ km s$^{-1}$ Mpc$^{-1}$. 
In this cosmology, the physical scale is 7.44 kpc arcsec$^{-1}$ at $z=3.34$.

\section{Observational data} \label{sec:observation}

\subsection{KCWI observations and data reduction} \label{sec:KCWIobsdata}

The right ascension and declination of the target LBG are 22:17:06.941 and $+$00:05:38.67,\footnote{The coordinate is based on the Subaru/Suprime-Cam $i$-band image, whose astrometry has been calibrated using stars in the Gaia DR2 catalogue \citep{2018A&A...616A...1G}. The declination is $0.''3$ north of the coordinate reported by \cite{2016ApJ...817..161M} and \cite{2023AJ....165..208M}, while the right ascension is quite similar to them.} in the SSA22 field \citep{1991ApJ...369...79L,1998ApJ...492..428S,2004AJ....128.2073H}. 
A DLA was found in the LBG spectrum previously obtained with the VLT/VIMOS \citep{2016ApJ...817..161M}. 
The redshift of the DLA was estimated at $z=3.335\pm0.007$ \citep{2016ApJ...817..161M} as a foreground cloud of the background LBG at $z=3.6061$ \citep{2023AJ....165..208M}. 
In this field, there are two other LBGs at $z=3.3366$ and $3.3082$, both identified by the Ly$\alpha$ emission line and a few interstellar metal absorption lines such as C~{\sc iv} \citep{2023AJ....165..208M}.
There is another galaxy detected in an emission line at 
106.3399 GHz (ALMA program ID\#2018.1.01427.S), which would be CO(4-3) at $z=3.33554$. 
This galaxy is not detected in $u,B,V,R_c,i',z',J,K$ and {\it Spitzer}/IRAC [3.6], but detected in [4.5], and is likely to be a molecular gas-rich submm galaxy (SMG) (Inoue et al. in preparation).

\begin{figure*}
\includegraphics[width=17cm]{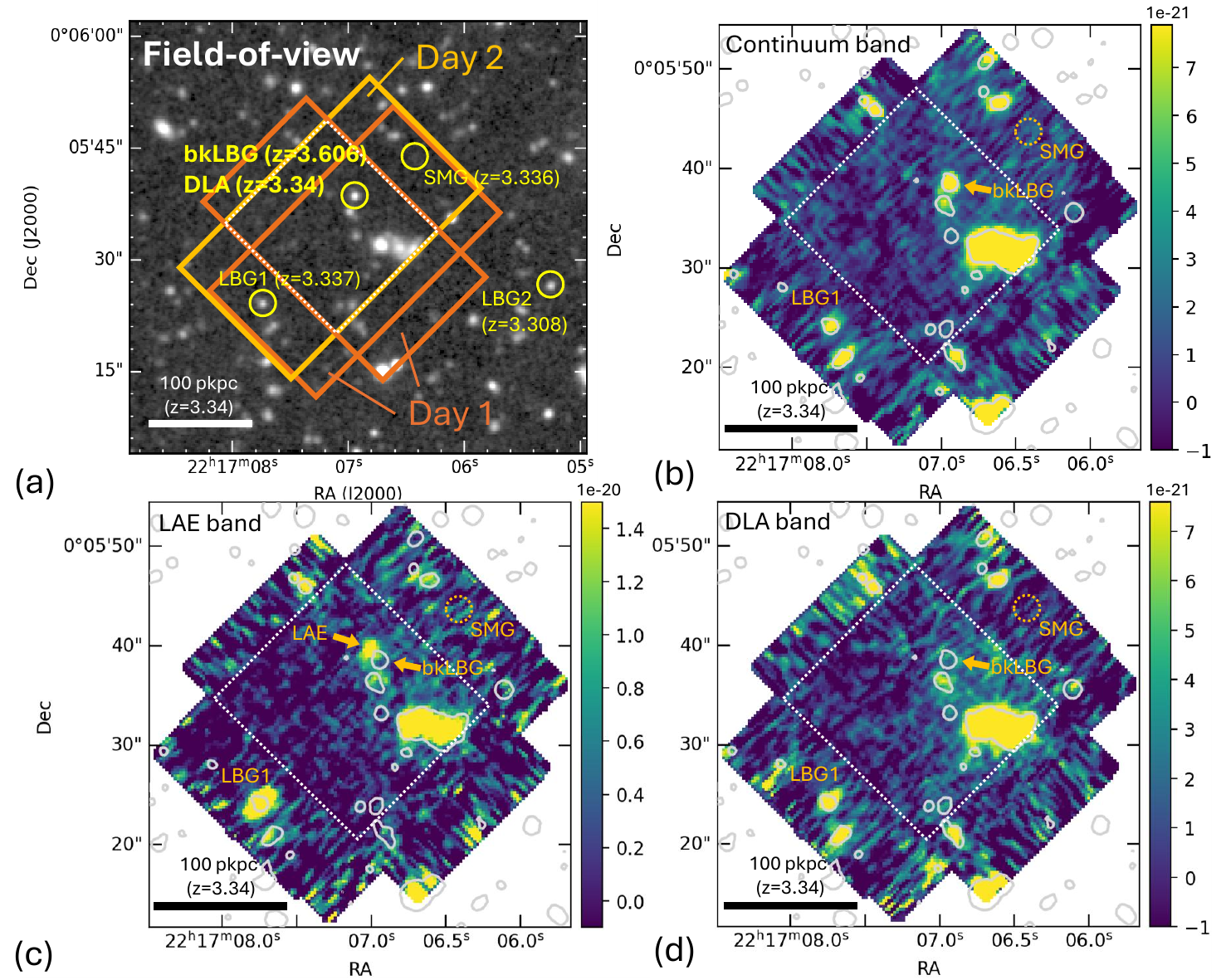}
\caption{(a) Configuration of the fields of view of the KCWI observations (dark and light orange rectangles for Day 1 and 2, respectively) superimposed on the Subaru/Suprime-Cam $i$-band image. The area where the on-source exposure time is greater than or equal to 19,200 seconds is defined as the deepest area, indicated by the white dotted line. The target DLA at $z=3.34$ is found in the spectrum of the background LBG (bkLBG) at $z=3.606$. There are two LBGs (LBG1 and LBG2) and one SMG with redshifts similar to the DLA as indicated. (b) The continuum image integrated over the wavelength ranges of [5019.0, 5068.5], [5434.0, 5458.5], and [5500.0, 5524.5] (in units of \AA) of the KCWI data cube. (c) The image integrated over a narrow wavelength range of [5276.5, 5285.5], showing an LAE very close to the DLA/bkLBG position. (d) The image integrated over the DLA wavelength range of [5250.0, 5276.0] and [5286.0, 5308.0], showing the very weak flux of bkLBG in this wavelength range. In the panels (b--d), the unit of the color bars is erg s$^{-1}$ cm$^{-2}$ \AA$^{-1}$ arcsec$^{-2}$, and the $1\sigma$ surface brightness levels are $1.2\times10^{-21}$, $3.4\times10^{-21}$, and $1.6\times10^{-21}$ in the unit, respectively. The contours show the $5\sigma$ surface brightness level in the Suprime-Cam $i$-band image and the dotted lines indicate the deepest area. The positions of bkLBG, LBG1, and SMG are also indicated.
\label{fig:FoVandImages}}
\end{figure*}

The observations (Keck Program ID: S369) were conducted on 13 and 14 July 2021 (UT), employing the KCWI \citep{2018ApJ...864...93M} installed on the Keck-II telescope of the Keck Observatory. 
The observations comprise three distinct fields-of-view (FOVs). 
Figure~\ref{fig:FoVandImages} (a) shows these FOVs and the positions of the target galaxy and galaxies at similar redshifts.
We used a set of Large-slicer and BM grating (spectral resolving power $R\sim2000$).
The total number of frames is 20 and each has 1,200 seconds on-source exposure. 
We define the deepest area where the total exposure time is greater than or equal to 19,200 seconds, which is shown by the yellow dotted rectangle in Figure~\ref{fig:FoVandImages} (a).
Tables~\ref{tab:instrument} and \ref{tab:observation} list basic information about the instrumental setup and the observations.

\begin{table}
 \caption{Instrumental setting of the KCWI observations.}
 \label{tab:instrument}
 \begin{tabular}{ll}
  \hline
  Image slicer & Large-slicer \\
  \hline
  Field-of-view & $20.''4\times33.''0$ \\
  Spatial pixel scale & $0.''29\times1.''35$ \\
  \hline
  Grating & BM grating\\
  \hline
  Spectral resolution & $R\sim2,000$ \\
  Wavelength range & 4,800--5,800 \AA\ \\
  Wavelength pixel scale & 0.5 \AA/pixel \\
  \hline
 \end{tabular}
\end{table}

\begin{table}
 \caption{A summary of Keck/KCWI observations.}
 \label{tab:observation}
 \begin{tabular}{llll}
  \hline
  Date & Position angle & Exposure & Nominal seeing \\
  \hline
  July 13, 2021 & $135^\circ$ & $5\times1,200$ sec & $0.''75$ \\
   & $45^\circ$ & $4\times1,200$ sec & \\
   \hline
  July 14, 2021 & $45^\circ$ & $11\times1,200$ sec & $0.''92$ \\
  \hline
 \end{tabular}
\end{table}

To create reduced data cubes from the 20 frames obtained during the observations, we used the Python version of the KCWI Data Reduction Pipeline (KCWIDRP) Version 1.0.\footnote{https://kcwi-drp.readthedocs.io/en/latest/versions.html\#version-1-0-2021}
We skipped the sky subtraction process in the KCWIDRP by modifying the program file kcwi\_pipeline.py.
This is because we did not want to subtract possible diffuse Ly$\alpha$ emission from H~{\sc i} gas around the DLA in the observing field, if there is any.
Next, we cut noisy spatial and wavelength regions near the edges of the data cubes, masked the strong atmospheric O~{\sc i} airglow at 5577.5--5587 \AA, and removed cosmic rays by performing $8\sigma$ clipping over the wavelength axis direction for each spatial pixel. 
We found a spatial gradient of the sky brightness in the data cubes. 
This also depends on the wavelength. 
We corrected the sky brightness gradient by adopting the method introduced in \cite{2022MNRAS.514.6074N}. 
Figure~\ref{fig:exampleshot} shows example images before and after correction.

Subsequently, we performed sky subtraction manually by the following way. 
First, we masked any possible continuum and emission line objects by conducting $2\sigma$ clipping twice along each spatial slice at each wavelength in the data cube. 
Second, we masked spatial regions ($8\times12$ pixels) potentially containing Ly$\alpha$ emission around the DLA. 
This procedure creates sky data cubes. 
We then performed sky subtraction by measuring the median values of the sky brightness at each wavelength in the sky data cubes and subtracting them from the original data cubes. 
Since the flattening of the sky background at each wavelength was achieved by the previous step described above, we can safely adopt a constant sky brightness in the subtraction.

The 20 sky-subtracted data cubes were then combined into three separate data sets for the three FOVs using the astronomical image mosaicking software \textsc{montage} \citep{2010ascl.soft10036J}.
\textsc{montage} stacks the data by reprojecting each onto new celestial coordinates based on the World Coordinate System (WCS). 
The new one pixel was set as $0.''2868\times0.''2868$. 
During the stacking process, the original pixel values are weighted by the area of overlap with the new pixels in the projection. 
The resulting area-weighted average determines the new pixel values. 
We have confirmed that the surface brightness in the celestial coordinate system is preserved in this procedure. 
For the reference of WCS, we used the Subaru/Suprime-Cam $i$-band image calibrated by the Gaia DR2 stars \citep{2018A&A...616A...1G}.

We performed flux calibration by using the data cubes of a standard star, Feige 110, which were obtained during observation and underwent the same processing as the target LBG data. 
We adopted the spectrum of Feige 110 taken from \cite{2014A&A...568A...9M} as the reference to derive the flux calibration factors as a function of the wavelength and applied them to the target data. 
Finally, we combined the flux-calibrated data cubes of the three FOVs into one data cube using the software \textsc{montage}. 
Figure~\ref{fig:FoVandImages} (b) shows a continuum image integrated over the wavelength ranges of [5019.0, 5068.5], [5434.0, 5458.5], and [5500.0, 5524.5] (in units of \AA).
It confirmed the WCS accuracy, as the objects found in the KCWI continuum image perfectly aligned with the positions shown by the contours of the Subaru/Suprime-cam $i$-band brightness. 
The root-mean-square (RMS) of the positional offsets between the KCWI and Suprime-Cam $i$-band images is $0.''14$, which is about a half pixel size of the final KCWI data cube. 
The point spread function (PSF) of the KCWI data cube was measured from the stacked data cube of the standard star, Feige 110.
The full width at half maximum (FWHM) of the PSF is $1.''572$, which is larger than the nominal seeing listed in Table~\ref{tab:observation}, likely due to the coarser spatial sampling originated from the slice width ($1.''35$).

\subsection{Additional spectroscopic data} \label{sec:additionalspectra}

The target LBG was previously observed with VLT/VIMOS and Keck/DEIMOS.
The VIMOS observations (ESO Program ID: 081.A-0081) were conducted in 2008 with the LR-Blue/OS-Blue setting that provides a spectral resolving power $R\sim180$.
The LBG was one of the 163 galaxies observed in that program.
The on-source exposure time was 14,080 seconds.
\cite{2019MNRAS.484.5868H} presents a full description of the target selection, observations, data reduction and redshift determination, while they are also briefly described in \cite{2016ApJ...817..161M}.

The DEIMOS observations (Keck Program IDs: S274D, S290D, S313D, and U066D) were conducted in 2015 and 2016 in the SSA22 H~{\sc i} Tomography Survey (SSA22-HIT; \citealt{2023AJ....165..208M}).
The 600ZD grating with a slit width of $1''$ was used, yielding a spectral resolving power $R\sim1000$.
The GG400 order blocking filter was used to cover a wide wavelength coverage of 4,000~\AA\ $< \lambda <$ 9,000~\AA.
The target LBG was included as one of the 78 observed galaxies in the Mask02, where the on-source exposure time was 7,200 seconds.
A full description of the target selection, observations, data reduction and redshift determination is presented in \cite{2023AJ....165..208M}.

\section{Spectral analysis}

In this section, we present a detailed spectral analysis of the background LBG.
Before a Voigt profile fit for the DLA in \S\ref{DLAvoigtfit}, we compare the KCWI spectrum with the previously obtained VIMOS and DEIMOS spectra in \S\ref{speccomparison} and report discoveries of a Ly$\alpha$ emitter (LAE) very close to the LBG/DLA position in \S\ref{laeidentification} and an O~{\sc i} $\lambda$1302 absorber on the line of sight to the LBG/DLA in \S\ref{oiabsorber}.

\begin{figure*}
\includegraphics[width=15cm]{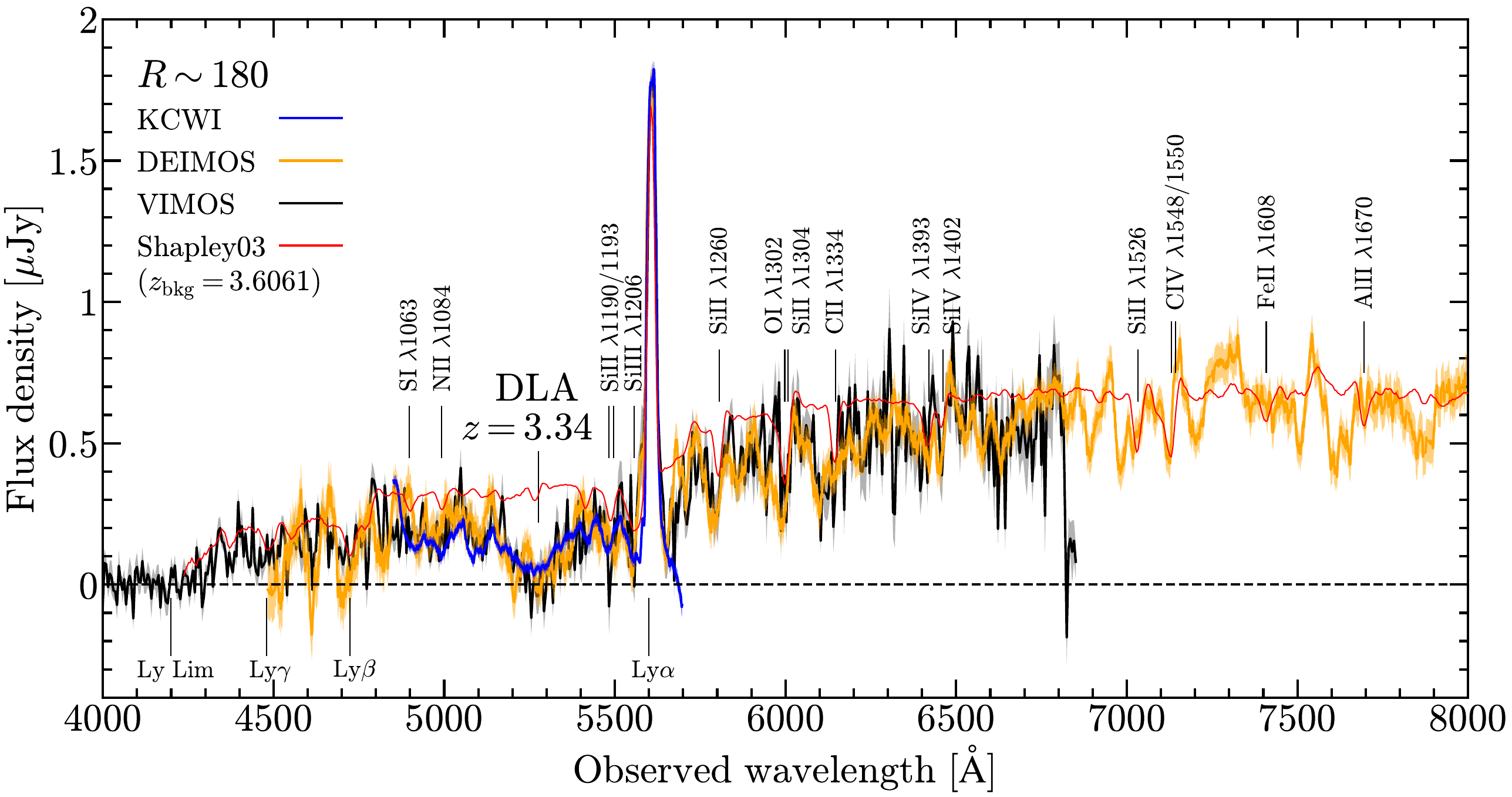}
\caption{A comparison of the background LBG spectra obtained with Keck/KCWI (blue; this work), Keck/DEIMOS (orange; \protect\citealt{2023AJ....165..208M}), and VLT/VIMOS (black; \protect\citealt{2016ApJ...817..161M}). The shaded range shows the $\pm1\sigma$ uncertainty of the spectra. The red curve is a typical LBG spectrum from \protect\cite{2003ApJ...588...65S} and redshifted to $z=3.6061$. The interstellar absorption features in the spectrum at $z=3.6061$ are noted. All spectra are boxcar-smoothed to a spectral resolution of $R(=\lambda/\Delta\lambda)\sim180$, the resolution of the VIMOS spectrum. The absolute flux densities are scaled to that of the VIMOS spectrum. The wavelength of the foreground DLA at $z=3.34$ is indicated.}
\label{fig:widerangespec}
\end{figure*}

\begin{figure*}
\includegraphics[width=15cm]{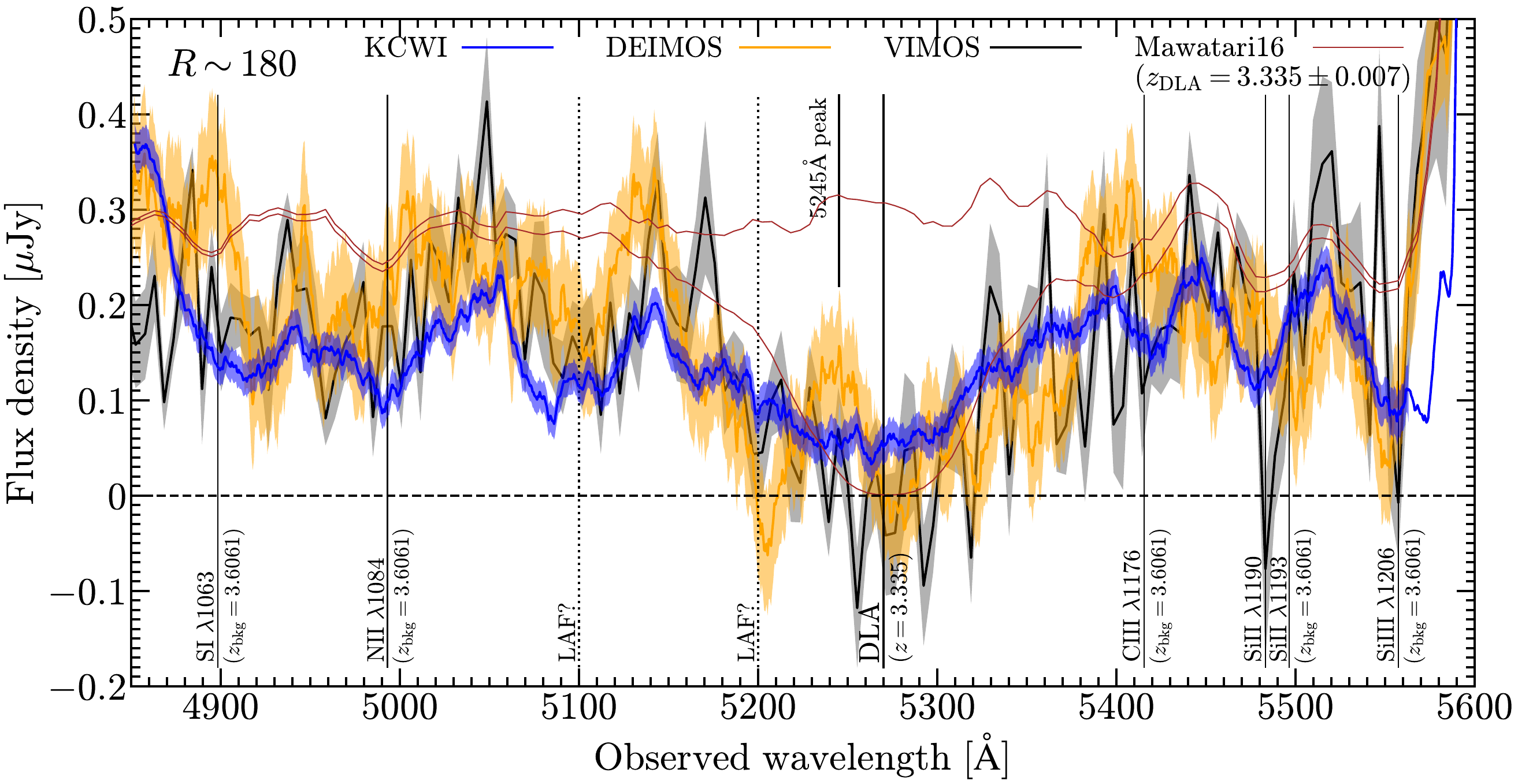}
\includegraphics[width=15cm]{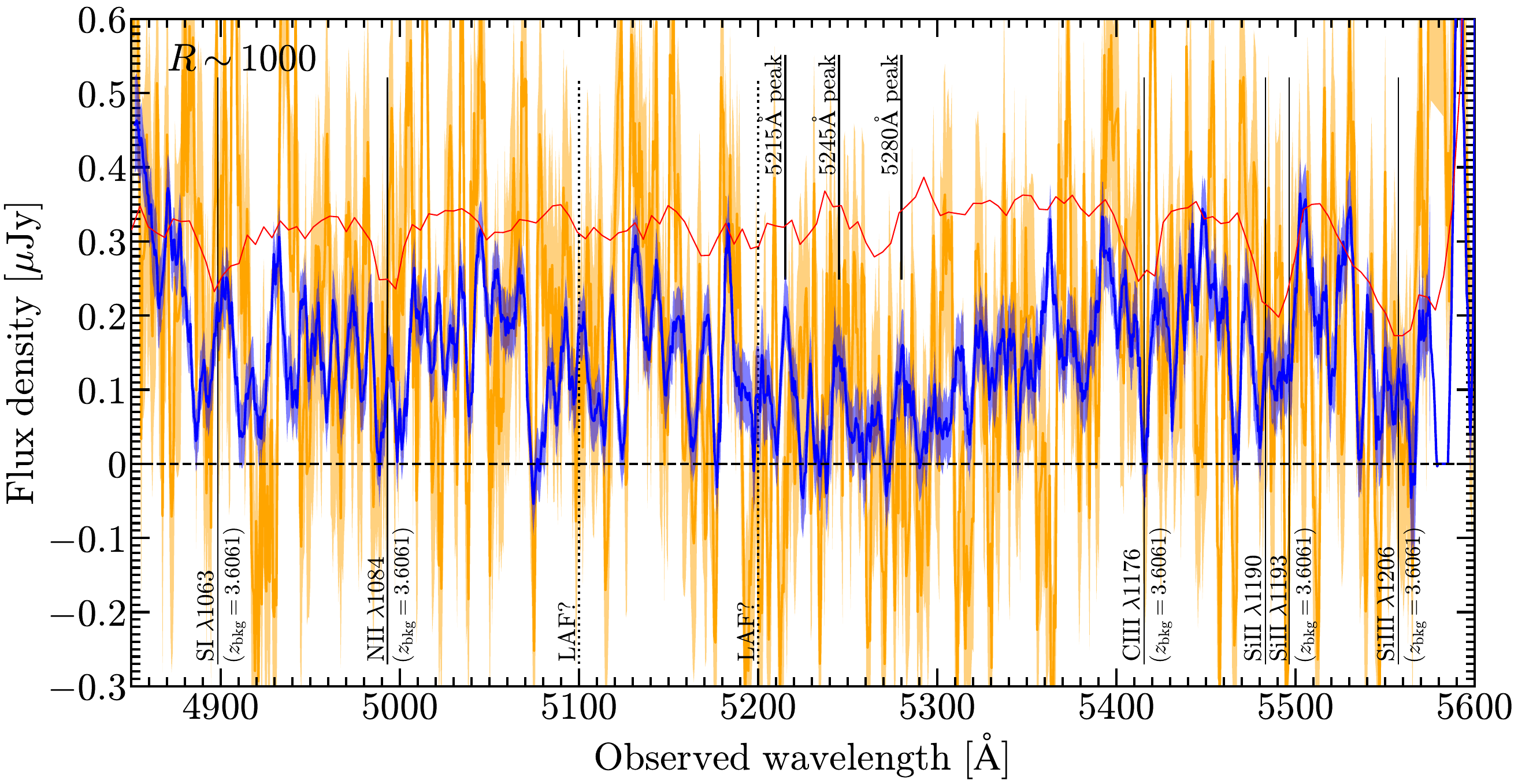}
\caption{Close-up views of the spectra around the DLA wavelength. (Top) The VIMOS spectral resolution ($R\sim180$) case. The brown curves are the LBG template spectra with/without the best-fit DLA profile from \protect\cite{2016ApJ...817..161M}. The vertical solid lines indicate absorption features in the background LBG spectrum and the DLA central wavelength at $z=3.335$. The vertical dotted lines are possible foreground Ly$\alpha$ forest (LAF) from \protect\cite{2016ApJ...817..161M}. A relatively high flux feature in the DLA range at 5245 \AA\ in the DEIMOS spectrum is also indicated. (Bottom) The DEIMOS spectral resolution ($R\sim1000$) case. Additional high flux features in the DLA range at 5215 \AA\ and 5280 \AA\ in the KCWI spectrum are indicated as well as the 5245 \AA\ feature.}
\label{fig:DLArangespec}
\end{figure*}

\subsection{Spectrum of the background galaxy and comparison with previous observations}
\label{speccomparison}

We extracted the background LBG spectrum in an aperture with a diameter of $1''.3$ from the final data cube. 
This choice of the extraction aperture is arbitrary and even smaller than the PSF FWHM, but any small change of the aperture does not affect the spectral features discussed throughout this paper. 
We also created an average sky spectrum from spectra extracted at random positions with the same diameter aperture in the deepest area of the final data cube. 
The average sky spectrum is consistent with zero at all wavelengths, while we found that a small but more positive trend at shorter wavelengths. 
Therefore, we fitted the average sky spectrum with a quadratic function at wavelengths below 5551.5 \AA\ and subtracted it from the LBG spectrum to remove the systematic trend likely caused by small residuals in the background subtraction.

As mentioned in \S\ref{sec:additionalspectra}, the LBG has been previously observed with the VLT/VIMOS \citep{2016ApJ...817..161M} and the Keck/DEIMOS \citep{2023AJ....165..208M}. 
We show a comparison of these spectra in Figures~\ref{fig:widerangespec} and \ref{fig:DLArangespec}.
The overall features, including the broad absorption over the wavelength range between $\sim5100$ \AA\ and $\sim5400$ \AA, are in excellent agreement, as shown in Figure~\ref{fig:widerangespec}.
However, as found in the top panel of Figure~\ref{fig:DLArangespec}, the smoothed ($R\sim180$) KCWI spectrum exhibits some residual flux around the central wavelength of the DLA, and the DEIMOS spectrum exhibits somewhat high flux around 5245 \AA\ compared to the VIMOS spectrum. 
In the case of $R\sim1000$ (the bottom panel of Figure~\ref{fig:DLArangespec}), both the KCWI and DEIMOS spectra show a peak around 5245 \AA, supporting the reality of this feature.
Furthermore, the KCWI spectrum shows two more peaks around 5215 \AA\ and 5280 \AA. 
It is likely that these features together produce the apparent residual flux in the smoothed ($R\sim180$) KCWI spectrum.
However, the DEIMOS spectrum does not show the peaks corresponding to the two additional peaks in the KCWI spectrum.
As we will discuss in \S\ref{laeidentification}, the 5280 \AA\ peak is caused by contamination from the Ly$\alpha$ emission line of a faint galaxy slightly northeast of the line of sight.
Since the slit position of the DEIMOS observation is east-west \citep{2023AJ....165..208M}, the Ly$\alpha$ emission is likely to be out of the DEIMOS slit.
However, it is unclear why the 5215 \AA\ peak does not appear in the DEIMOS spectrum.
In the KCWI cube data we could not find any corresponding Ly$\alpha$ emission for the 5215 \AA\ and 5245 \AA\ peaks (see Figures~\ref{fig:5215Apeak} and \ref{fig:5245Apeak}), suggesting that they are transmission peaks newly found in the higher resolution ($R\gtrsim1000$) spectrum from KCWI (and also DEIMOS for the 5245 \AA\ peak).
If this is the case, the DLA identified in the low resolution ($R\sim180$) VIMOS spectrum may actually be composed of multiple absorption systems (see a discussion in \S\ref{DLAvoigtfit}).

\subsection{Identification of Ly\texorpdfstring{$\alpha$}{} emission close to the line of sight}
\label{laeidentification}

\begin{figure}
\includegraphics[width=\columnwidth]{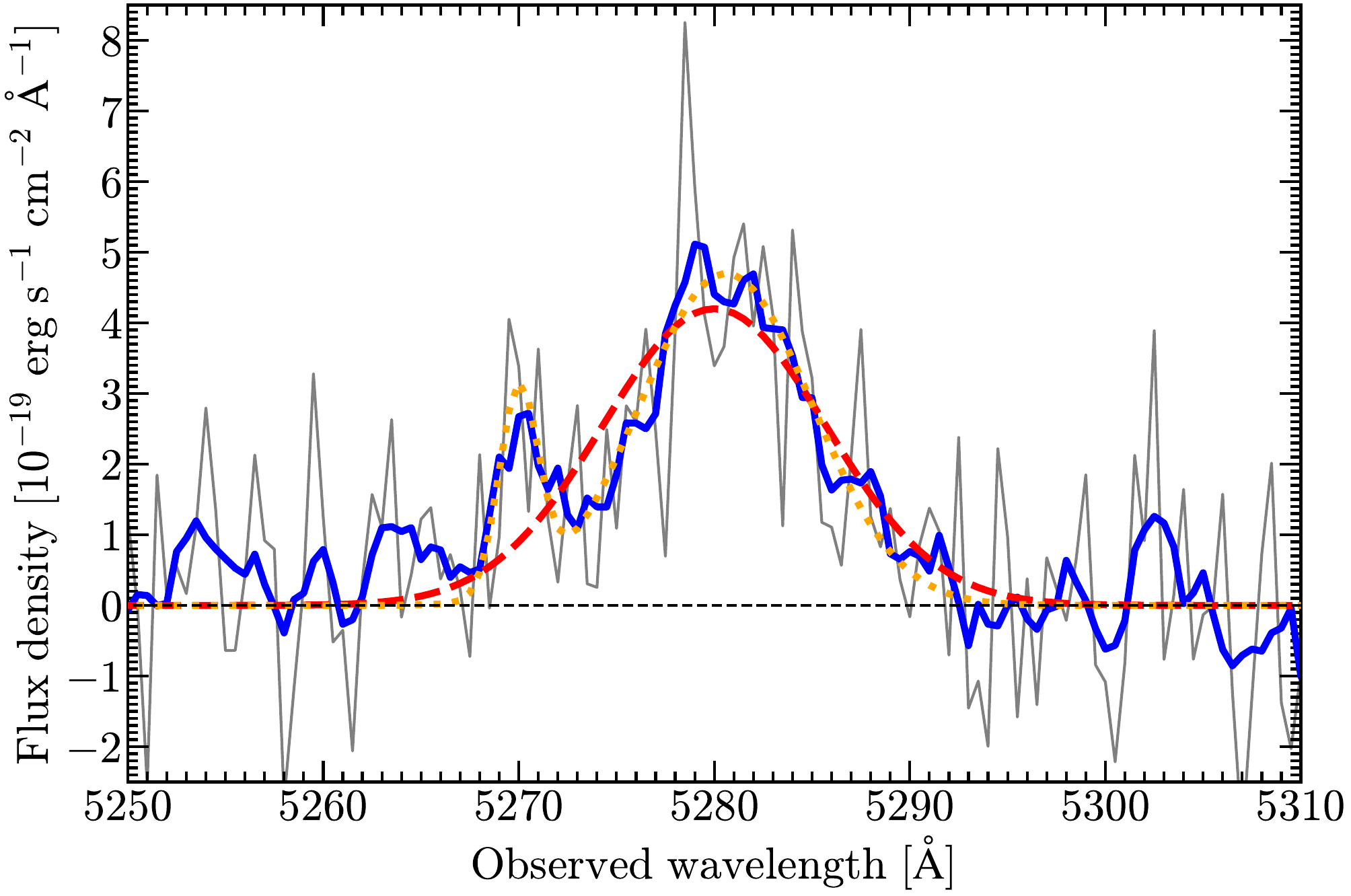}
\caption{LAE spectrum extracted from the KCWI data cube within an elliptical aperture of $2''.7\times0''.9$ around the position of the LAE close to the DLA (Figure~\ref{fig:FoVandImages}c). The grey curve is the original spectrum, while the blue curve is a 5-pixel boxcar smoothed spectrum. The red dashed curve is the best-fit single Gaussian function, while the orange dotted curve is the case with a double Gaussian function.
\label{fig:LAEspec}}
\end{figure}

The 5280 \AA\ feature found in the KCWI spectrum (the bottom panel of Figure~\ref{fig:DLArangespec}) is contamination from an emission line from an object slightly offset to the northeast from the LBG line of sight as shown in Figure~\ref{fig:FoVandImages} (c). 
The integration wavelength range [5276.5, 5285.5] was chosen to be approximately one FWHM of the emission line profile.
The spatial position of the emission line well aligns with the marginal subcomponent in the Suprime-Cam images reported by \cite{2016ApJ...817..161M} (see their Fig.~3). 
The emission line spectrum extracted with an elliptical aperture covering its spatial extent is shown in Figure~\ref{fig:LAEspec}.
If this emission line is Ly$\alpha$, the redshift is $z=3.3433\pm0.0005$ from a least-squares fit with a Gaussian function to the emission line profile.
The line velocity FWHM is $757\pm88$ km s$^{-1}$, which is broad enough to neglect the line spread function effect ($R\sim2000$, corresponding to $\Delta v\sim150$ km s$^{-1}$) and one of the largest values found in the literature \citep[e.g.,][]{2013ApJ...765...70H}.
A closer look at the spectrum may suggest the presence of a blueshifted component.
A double Gaussian function fit gives a reasonably good result with FWHMs of $\approx130$ km s$^{-1}$ and $\approx590$ km s$^{-1}$ for the blue and red components respectively.
The spectral minimum is located at 5273 \AA, corresponding to $z\approx3.338$.
Table~\ref{tab:laeproperties} is a summary of the measurements.
The impact parameter $b$ between the centroids of the LAE and the DLA is $b=1''.73\pm0''.28$. 

Other features at 5215 \AA\ and 5245 \AA\ found in the bottom panel of Figure~\ref{fig:DLArangespec} do not show any emission line object in the KCWI cube 
(see Appendix B).
Therefore, we consider these features to be transmission peaks in the LBG spectrum.

\begin{table}
 \caption{Properties of the Ly$\alpha$ emitter.}
 \label{tab:laeproperties}
 \begin{tabular}{lll}
  \hline
  Property & Measurement & Remark \\
  \hline
  Redshift & $3.3433\pm0.0005$ & Single Gaussian \\
  & $\approx3.338$ & Double Gaussian \\
  FWHM (km s$^{-1}$) & $757\pm88$ & Single Gaussian \\
  & $\approx130$, $590$ & Double Gaussian \\
  Ly$\alpha$ flux ($10^{-18}$ erg s$^{-1}$ cm$^{-2}$) & $4.78\pm1.27$ & \\
  Ly$\alpha$ luminosity ($10^{41}$ erg s$^{-1}$) & $4.79\pm1.28$ & \\
  Ly$\alpha$ equivalent width (\AA) & $>12.3$ ($3\sigma$) & Rest-frame \\
  SFR(Ly$\alpha$) (M$_\odot$ yr$^{-1}$) & $0.44\pm0.12$ & Strict lower bound \\
  \hline
 \end{tabular}
\end{table}

\subsection{Identification of O~{\sc i} \texorpdfstring{$\lambda1302$}{} absorption}
\label{oiabsorber}

\begin{figure}
\includegraphics[width=\columnwidth]{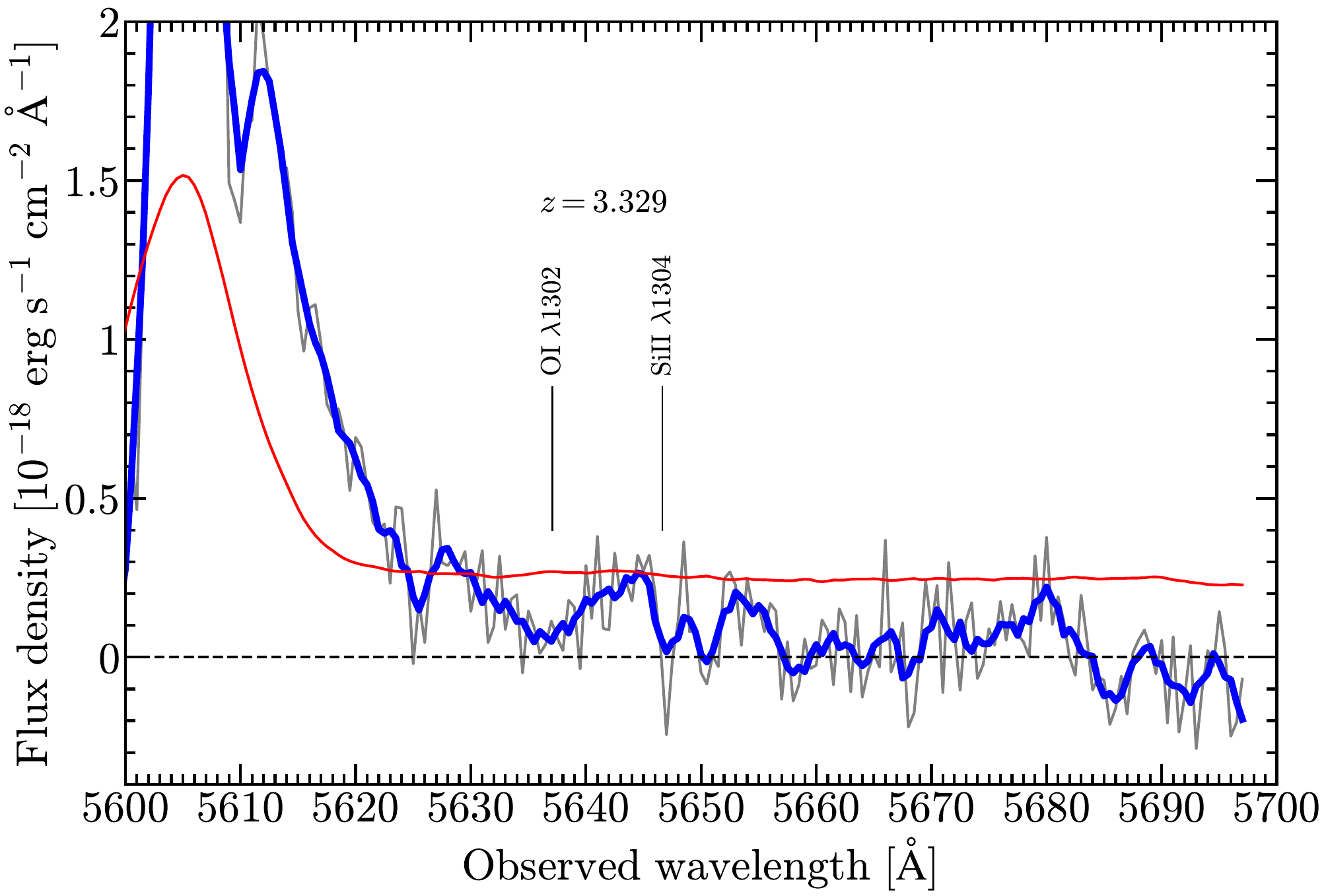}
\includegraphics[width=\columnwidth]{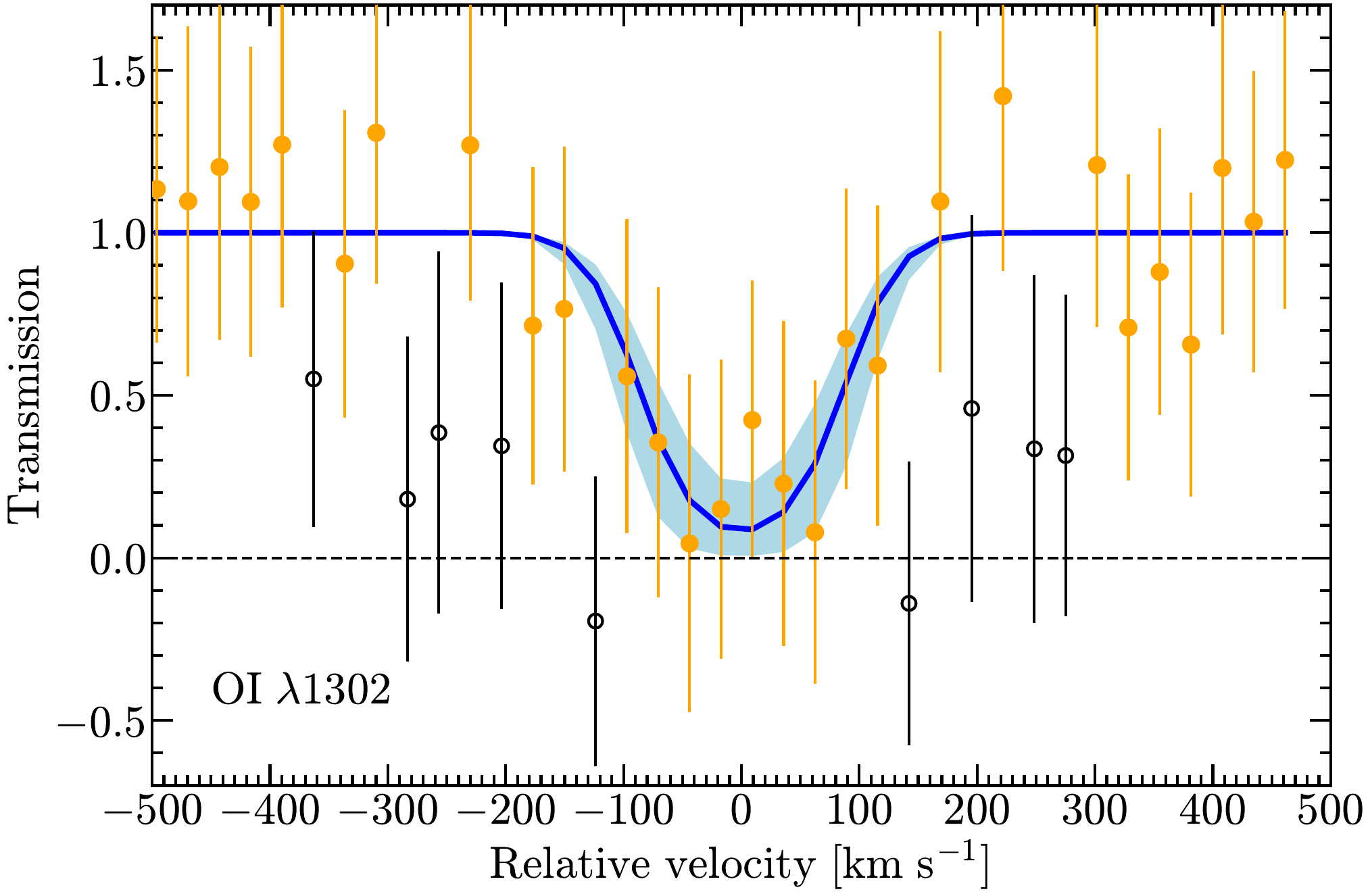}
\caption{O~{\sc i} $\lambda 1302$ absorption. (Top) Part of the KCWI spectrum of the background LBG, wavelengths longer than its Ly$\alpha$ line. The grey curve is the original spectrum, while the blue curve is a 5-pixel boxcar smoothed spectrum. The red curve is the best-fit continuum spectrum obtained in \S\ref{DLAvoigtfit}. Although the Ly$\alpha$ line of the best-fit spectrum is much weaker than the observed one, it does not affect any results in this paper because we do not discuss the Ly$\alpha$ emission of the background LBG. An absorption feature around 5637 \AA\ is attributed to the O~{\sc i} $\lambda 1302$ absorption at $z=3.329$. There could be another absorption, possibly Si~{\sc ii} $\lambda 1304$ at the same redshift. However, the spectrum at wavelengths longer than $\sim5650$ \AA\ may be affected by the lower sensitivity due to the edge of coverage. (Bottom) The transmission spectrum, which is the spectrum normalized by the best-fit continuum. The wavelength range is 5627.5 and 5645.5 \AA, which is converted to the velocity relative to the best-fit systemic velocity at $z=3.3288$. The orange and black data points with error bars are, respectively, the data used and not used for Gaussian fitting to the absorption feature. The blue line is the best-fit result, and the cyan region shows its $\pm1\sigma$ uncertainty range.
\label{fig:OI1302}}
\end{figure}

In the KCWI spectrum shown in the top panel of Figure~\ref{fig:OI1302}, we have found an absorption feature around the wavelength of $\sim5637$ \AA, identified as the O~{\sc i} $\lambda1302$ line at $z\sim3.33$.
This absorption feature was also suggested in the previous VIMOS spectrum (see Fig.~1 in \citealt{2016ApJ...817..161M}). 
There may be another absorption feature around the wavelength of $\sim5647$ \AA, which could be the Si~{\sc ii} $\lambda1304$ line at the same redshift as O~{\sc i}.
However, the wavelength is close to the edge of coverage and sensitivity may be systematically lower.
In fact, at wavelengths longer than $\sim5650$ \AA, the spectrum becomes distributed around zero.
Therefore, we only discuss the O~{\sc i} $\lambda1302$ absorption in this paper.

The bottom panel of Figure~\ref{fig:OI1302} shows the transmission spectrum, which is the spectrum normalized by the best-fit LBG continuum spectrum (see \S\ref{DLAvoigtfit}).
The horizontal axis shows the velocity centered around the wavelength of the O~{\sc i} $\lambda1302$ absorption. 
We performed a least-squares fit to the absorption feature, assuming that the optical depth follows a Gaussian function with its standard deviation equal to the spectral resolution of the KCWI observation, $\sim5000/2000/2\sqrt{2ln2} \approx 1.0$ \AA\ in the observed frame. 
We did not consider the covering fraction in this analysis for simplicity (but see for the DLA Voigt profile analysis in \S\ref{DLAvoigtfit}). 
We selected the data points shown in orange in Figure~\ref{fig:OI1302} to be used for the profile fitting, removing some outlier data points shown in black that are about 1$\sigma$ or more lower than the continuum level and the absorption feature.
The obtained best-fit with its $\pm1\sigma$ uncertainty is shown in blue and cyan in the bottom panel of Figure~\ref{fig:OI1302}, respectively. 
The best-fit redshift does not change even if we use all data points in the wavelength range, but the column density, discussed below, becomes a few times larger. 
Table~\ref{tab:OI1302properties} is a summary of the measured quantities. 

The redshift of the O~{\sc i} absorption is well determined at $z=3.3288\pm0.0004$ thanks to the narrow absorption feature. 
The equivalent width was calculated by integrating the transmission spectrum of the best-fit function. 
We also estimated the O~{\sc i} column density by integrating the optical depth spectrum of the best-fit function based on the Apparent Optical Depth Method (AODM; \citealt{1991ApJ...379..245S}). 
Given that the absorption feature reaches near zero flux, the O~{\sc i} line is most likely to be saturated. 
Therefore, the estimated O~{\sc i} column density should be regarded as a strict lower bound (e.g., \citealt{1999ApJS..121..369P}). 

We also tried a Voigt function fit to the absorption feature. 
However, the column density and the Doppler parameter are strongly degenerated due to insufficient spectral resolution. 
Assuming a Doppler parameter of 75 km s$^{-1}$, similar to the KCWI spectral resolution, we found the O~{\sc i} column density to be similar to that from the Gaussian fitting, but 0.3-dex higher. 
As the Doppler parameter decreases, the O~{\sc i} column density increases. 
We need much higher spectral resolution to estimate the O~{\sc i} column density more accurately.

\begin{table}
 \caption{Properties of O~{\sc i} $\lambda1302$ absorption.}
 \label{tab:OI1302properties}
 \begin{tabular}{lll}
  \hline
  Property & Measurement & Remark \\
  \hline
  Redshift & $3.3288\pm0.0004$ & \\
  Equivalent width (\AA) & $0.72^{+0.22}_{-0.17}$ & Rest-frame \\
  log$_{10}$($N_{\rm OI}$/cm$^{-2}$) & $15.30^{+0.32}_{-0.22}$ & Strict lower bound \\
  \hline
 \end{tabular}
\end{table}

\subsection{Voigt profile fit to the broad absorption feature and confirmation of the DLA}
\label{DLAvoigtfit}

\begin{figure*}
\includegraphics[width=15cm]{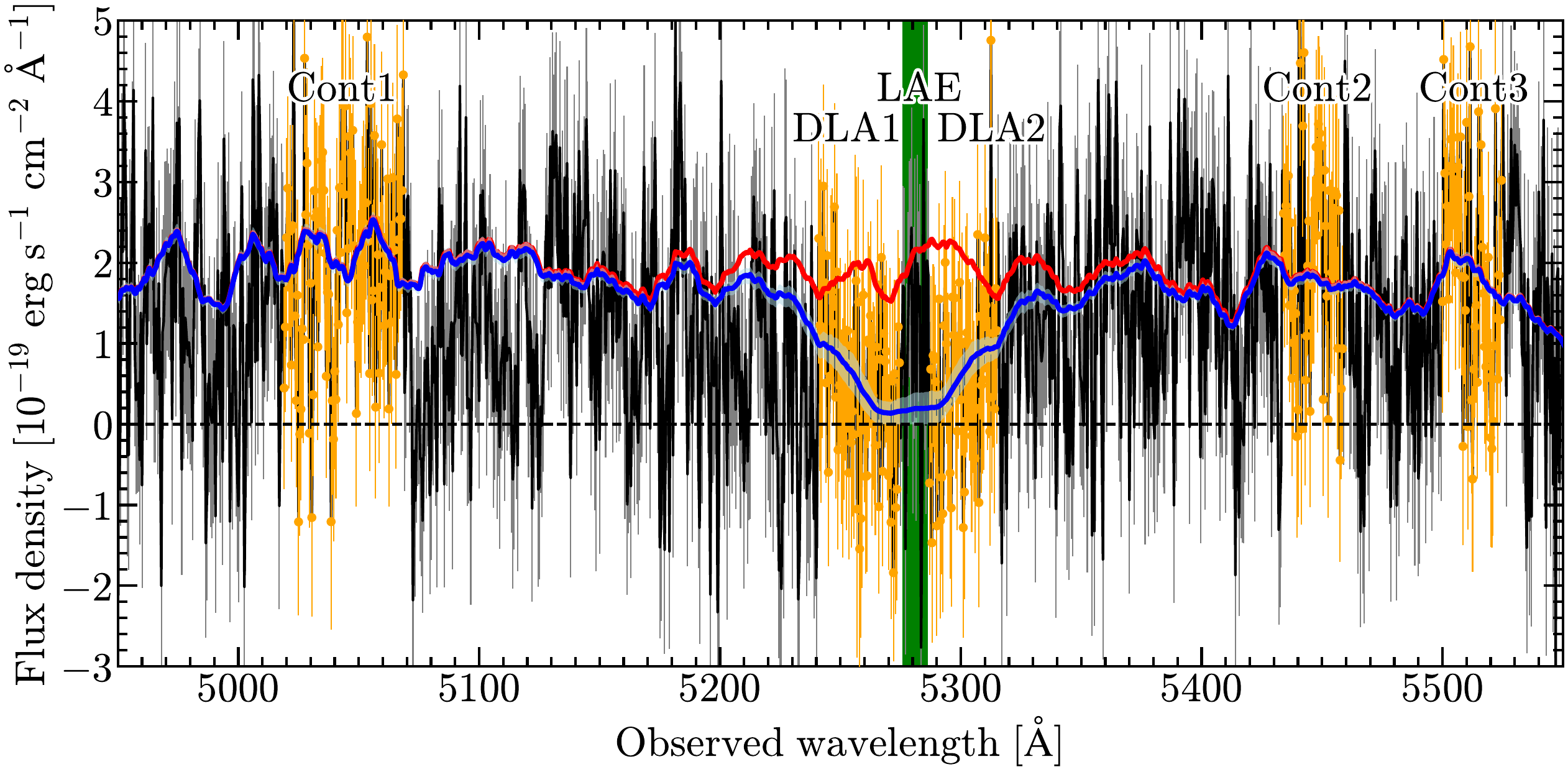}
\caption{The result of the Voigt profile fit to the broad absorption feature in the background LBG spectrum extracted from the KCWI data cube (black line, $\pm1\sigma$ uncertainty is shown by the thin line, without any smoothing). The orange points with the error bars are the data used for the fit (see Table~\ref{tab:wavelengths}), selected to avoid the wavelengths of possible foreground Ly$\alpha$ forest and metal absorption lines of the background LBG as well as the Ly$\alpha$ line from the LAE close to the DLA (the green hatched range). The red line shows the best fit LBG continuum spectrum, while the blue line is the best-fit spectrum with a DLA. The cyan range shows its $\pm1\sigma$ uncertainty.
\label{fig:Voigtfit}}
\end{figure*}

\begin{figure}
\includegraphics[width=\columnwidth]{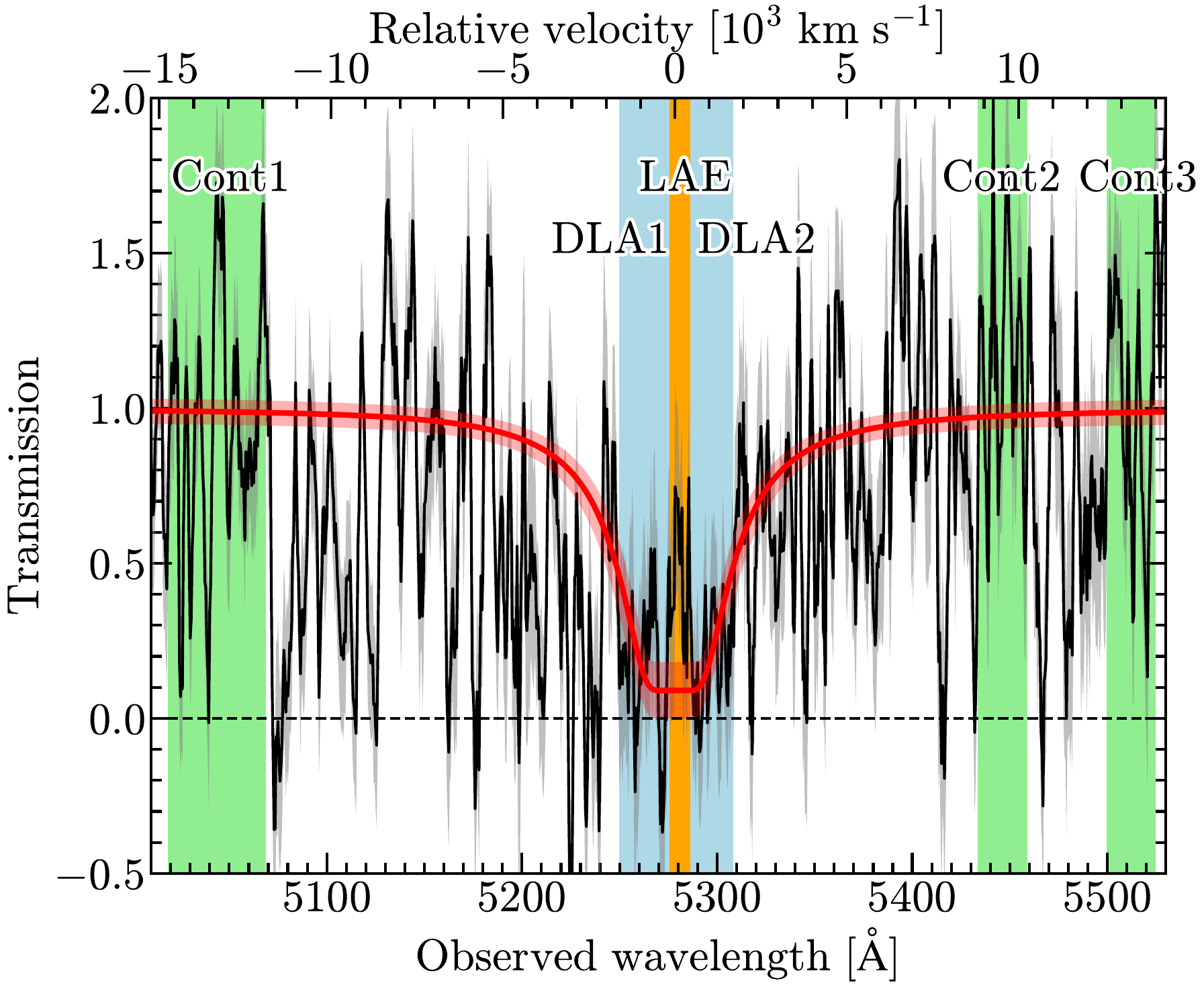}
\caption{The DLA transmission. The black solid line shows the transmission spectrum (i.e. the observed spectrum divided by the best-fit continuum spectrum). A boxcar-smoothing of a 2.5 \AA\ width (5 spectral pixels) has been applied. The grey range shows the range of $\pm1\sigma$ uncertainty. The red curve with the shaded range is the best-fit DLA profile with its $\pm1\sigma$ uncertainty. The coloured wavelength ranges show the ranges to create the images of the continuum (green; Figure~\ref{fig:FoVandImages} [b]), the Ly$\alpha$ emitter (orange; Figure~\ref{fig:FoVandImages} [c]), and the DLA (blue; Figure~\ref{fig:FoVandImages} [d]). The wavelength ranges are listed in Table~\ref{tab:wavelengths}. The top horizontal axis shows the radial velocity relative to the best-fit redshift of $z=3.3419$.
\label{fig:DLAtransmission}}
\end{figure}

\begin{figure}
\includegraphics[width=\columnwidth]{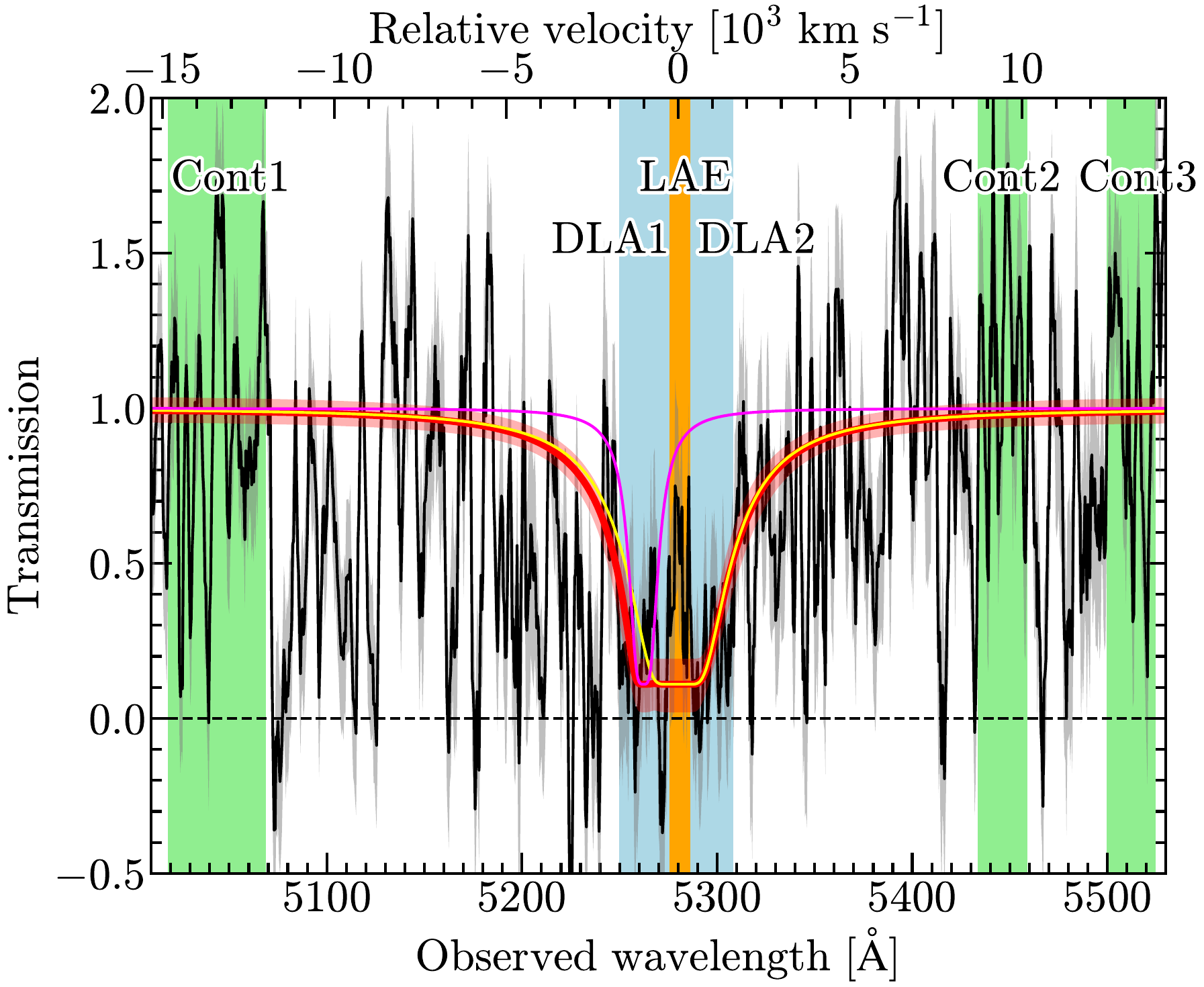}
\caption{Same as Figure~\ref{fig:DLAtransmission} but for the two-component fit. The velocity coordinate in the top horizontal axis is relative to the LAE redshift of $z=3.3433$. The thin curves in magenta and yellow show the transmission of each absorber of the two components.
\label{fig:2comp}}
\end{figure}

\begin{table}
 \caption{Selected wavelength ranges (\AA).}
 \label{tab:wavelengths}
 \begin{tabular}{lll}
  \hline
   & Voigt profile fit & Narrowband image \\
  \hline
  Continuum 1 & 5019.0--5068.5 & 5019.0--5068.5 \\
  DLA 1 & 5241.0--5274.5 & 5250.0--5276.0 \\
  LAE & --- & 5276.5--5285.5 \\
  DLA 2 & 5287.0--5315.0 & 5286.0--5308.0 \\
  Continuum 2 & 5434.0--5458.5 & 5434.0--5458.5 \\
  Continuum 3 & 5500.0--5524.5 & 5500.0--5524.5 \\
  \hline
 \end{tabular}
\end{table}

\begin{table*}
 \caption{Properties of the DLA. The single component fit is the case shown in Figures~\ref{fig:Voigtfit} and \ref{fig:DLAtransmission}, where the redshift is a free parameter as well as the H~{\sc i} column density and the covering fraction. The two-component fit is the case shown in Figure~\ref{fig:2comp}, where we have assumed two absorbers at the fixed redshifts of the LAE and the O~{\sc i} absorber. The two absorbers have different H~{\sc i} column density but the same covering fraction.}
 \label{tab:DLAproperties}
 \begin{tabular}{llll}
  \hline
  Property & 1 component & 2 components & \cite{2016ApJ...817..161M} \\
  \hline
  Redshift & $3.3419\pm0.0025$ & 3.3433 (LAE), 3.3288 (O~{\sc i}) & $3.335\pm0.007$ \\
  log$_{10}$($N_{\rm HI}$/cm$^{-2}$) & $20.96\pm0.15$ & $20.88\pm0.17$ (LAE), $19.77_{-1.77}^{+0.43}$ (O~{\sc i}) & $21.86\pm0.17$ \\
  Covering fraction & $0.91\pm0.09$ & $0.89\pm0.09$ & $>0.7$ (2$\sigma$) \\
  \hline
 \end{tabular}
\end{table*}

We have performed extensive Voigt profile fitting to the broad absorption feature in the KCWI spectrum of the original spectral resolution of $R\sim2000$ without smoothing by adopting a least-squares fit. 
The transmission peak around 5245 \AA\ discussed in \S\ref{speccomparison} is considered to be part of the damping wing of the DLA.
The wavelength range contaminated by the LAE is removed from the fitting.
The continuum regions were chosen to be sufficiently far from the DLA absorption.
Table~\ref{tab:wavelengths} is a summary of the wavelength ranges.

We consider two cases: a single absorber and a set of two absorbers.
While in the first case the redshift of the single absorber is a free parameter, in the second case we fixed the redshifts of the two absorbers to be the same as those of the LAE and the O~{\sc i} absorber (\S\ref{laeidentification} and \ref{oiabsorber}).
Other fitting parameters are the H~{\sc i} column density, the covering fraction, and the continuum level.
The covering fraction was not included in the analysis of \cite{2016ApJ...817..161M} but is newly adopted to evaluate possible residual flux at the bottom of the absorption feature. 
For the single absorber case, the observed spectrum is given by $F_\lambda^{\rm obs}=(1-f_{\rm cov}) F_\lambda^{\rm cont}+f_{\rm cov} T_\lambda^{\rm DLA} F_\lambda^{\rm cont}$, where $f_{\rm cov}$, $T_\lambda^{\rm DLA}$, and $F_\lambda^{\rm cont}$ are the covering fraction, the DLA transmission, and the continuum spectrum, respectively. 
The first term of the equation is the residual flux, while the second term is the transmitted flux through the DLA.
For the two-component absorber case, we assumed for simplicity that the two absorbers spatially perfectly overlap and thus we adopted a single value of $f_{\rm cov}$.\footnote{For a general case with two absorbers \citep[e.g.,][]{2021ApJ...921..119I}, we can express $F_\lambda^{\rm obs}=(1-f_{\rm cov1}-f_{\rm cov2}+p)F_\lambda^{\rm cont}+(f_{\rm cov1}-p)T_\lambda^{\rm abs1}F_\lambda^{\rm cont}+(f_{\rm cov2}-p)T_\lambda^{\rm abs2}F_\lambda^{\rm cont}+pT_\lambda^{\rm abs1}T_\lambda^{\rm abs2}F_\lambda^{\rm cont}$, where the terms with `1' and `2' indicate quantities of the first and second absorbers, respectively, and $p$ ($0\leq p \leq \min(f_{\rm cov1},f_{\rm cov2})$) is the areal fraction where the two absorbers overlap in front of the background source. For the perfect overlap case, $p=f_{\rm cov1}=f_{\rm cov2}$, and we denote $p=f_{\rm cov}$ and $T_\lambda^{\rm abs1}T_\lambda^{\rm abs2}=T_\lambda^{\rm DLA}$.}
For the continuum template in fitting, we adopted an average spectrum of 113 LBGs with a clear Ly$\alpha$ emission line in the SSA22 field observed with Keck/DEIMOS (the ‘Ae’ sample in \citealt{2023AJ....165..208M}). 
We note that the continuum spectrum, $F_\lambda^{\rm cont}$, includes the effect of mean Ly$\alpha$ forest (LAF). 
Namely, $F_\lambda^{\rm cont}=F_\lambda^{\rm cont,int} \times T_\lambda^{\rm LAF}$, where $F_\lambda^{\rm cont,int}$ and $T_\lambda^{\rm LAF}$ are the intrinsic continuum spectrum without LAF and the transmission through LAF, respectively. 
The mean redshift of the LBG sample of the template spectrum is $z=3.455$, slightly smaller than but similar to the background LBG of $z=3.6061$, and we neglected the small difference of the mean LAF transmission between these two redshifts. 
For the Voigt function, we assumed a Doppler parameter of 25 km s$^{-1}$, while the results did not change even in the case of 100 km s$^{-1}$. 

Figures~\ref{fig:Voigtfit} and \ref{fig:DLAtransmission} show the result of the single absorber case and Figure~\ref{fig:2comp} shows the result of the two-component absorber case.
The obtained DLA properties are listed in Table~\ref{tab:DLAproperties} and compared with those in \cite{2016ApJ...817..161M}. 
Although the redshift is consistent with that of \cite{2016ApJ...817..161M}, the H~{\sc i} column density is an order of magnitude smaller than theirs because the higher spectral resolution of the KCWI spectrum suggests that the broad H~{\sc i} absorption is composed of multiple absorption systems.
For example, the low transmission around 5230 \AA\ and 5260 \AA\ is now a different H~{\sc i} absorption component from the DLA, whereas it was part of the DLA in the low resolution VIMOS spectrum (see the top panel of Figure~\ref{fig:DLArangespec}).
In the two-component case, the sum of the H~{\sc i} column densities of the two absorbers is consistent with the single-component case, although the uncertainty of the weaker component corresponding to the O~{\sc i} absorber is large.

\section{The DLA silhouette image}

\subsection{Making the DLA silhouette image}

We created a ``DLA silhouette'' (i.e. the negative brightness) image by subtracting the continuum image (Figure~\ref{fig:FoVandImages} [a]) of the background LBG from the DLA narrowband image (Figure~\ref{fig:FoVandImages} [d]). 
The wavelength ranges of these images are listed in Table~\ref{tab:wavelengths} and visually shown in Figures~\ref{fig:DLAtransmission} and \ref{fig:2comp}.
Note that the DLA wavelength ranges are narrower than those used for the Voigt profile fit to avoid slightly higher transmission wavelengths in the damping wing parts.

\begin{figure*}
\includegraphics[width=13cm]{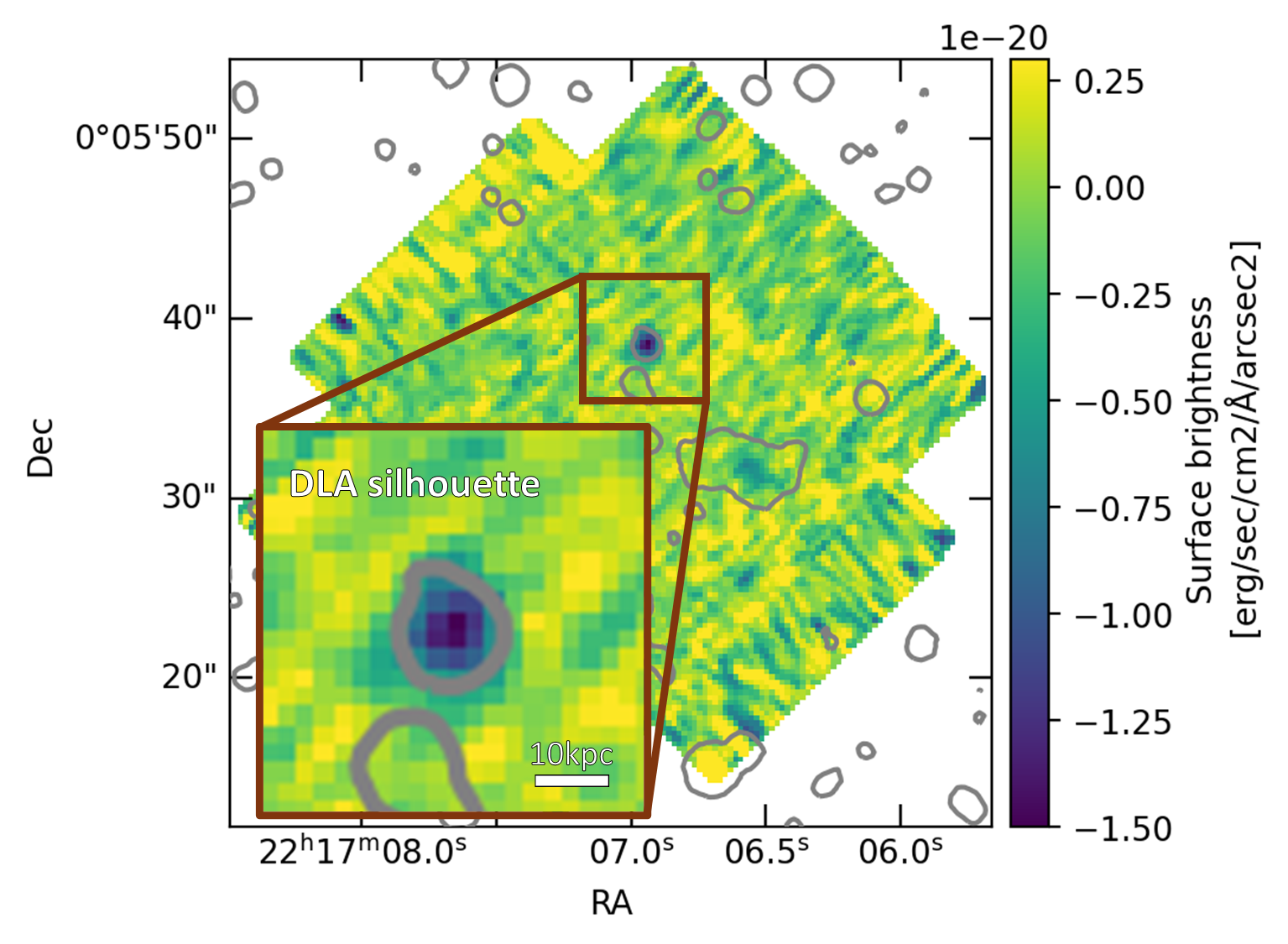}
\caption{The DLA ``silhouette'' image produced by subtracting the continuum image from the narrowband image of the DLA wavelength range. The negative surface brightness is caused by the DLA imprinted in the background LBG continuum. The gray contours are the $5\sigma$ surface brightness level in the Subaru/Suprim-Cam $i$-band image, indicating the positions and shapes of the continuum sources.
\label{fig:DLAsilhouette}}
\end{figure*}

The result is presented in Figure~\ref{fig:DLAsilhouette}, showing the clear negative ``hole'' at the position of the background LBG.
This is the ``silhouette'' produced by the DLA: the H~{\sc i} absorption against the background LBG continuum.\footnote{Even if we include the damping wing parts, the silhouette shape does not change because the background LBG is not resolved.}
We measured the signal-to-noise ratio of the silhouette by performing aperture photometry with a diameter of $2.''0$.
The uncertainty of the silhouette was measured by random aperture photometry with the same aperture in the deepest area (exposure time $\geq19,200$ seconds), excluding the areas around the DLA (i.e. the background LBG) and the southern bright foreground galaxies. 
The results of the random aperture photometry follow a Gaussian function around zero, ensuring excellent sky and continuum subtraction. 
The obtained signal-to-noise ratio is 8.8.

\subsection{Radial profile of the DLA silhouette}
\label{sec:radialprofile}

\begin{figure*}
\includegraphics[width=6cm]{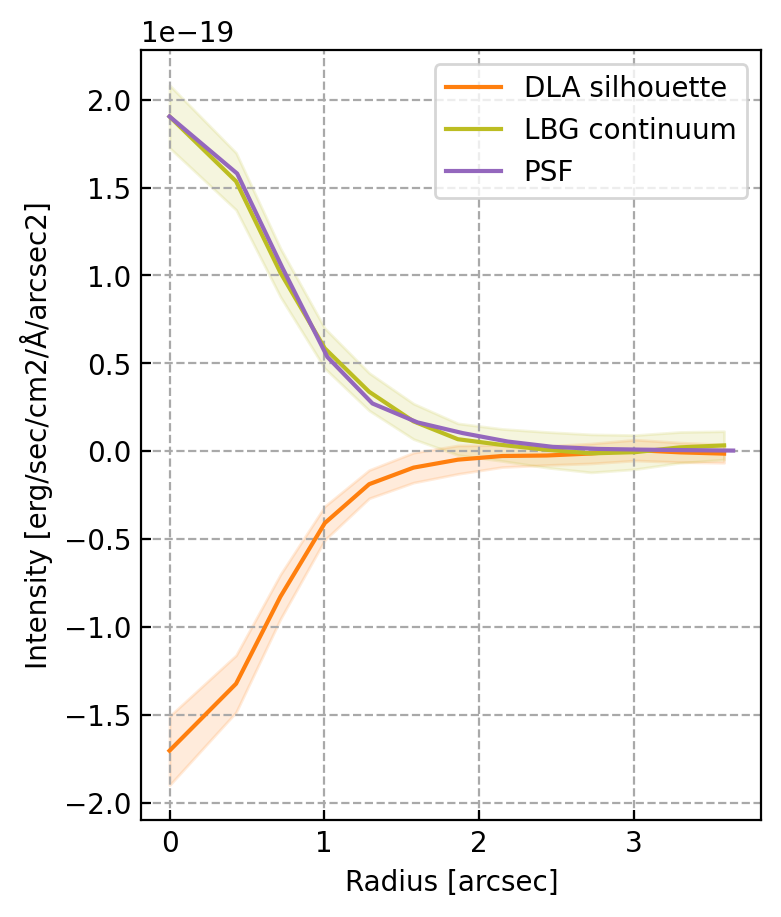}
\includegraphics[width=6cm]{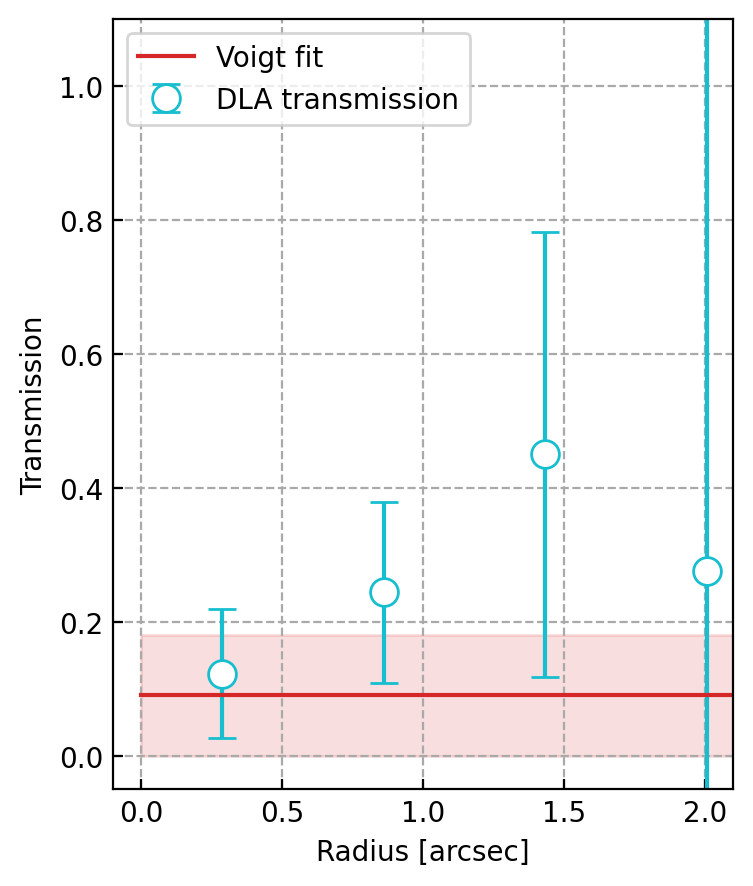}
\caption{The radial profiles of the DLA ``silhouette'', the background LBG continuum, and the point spread function (PSF) (left) and the DLA transmission (right). The PSF profile is scaled to the central intensity of the LBG profile. The DLA transmission is compared to that estimated from the Voigt function fit to the spectral profile of the DLA (see \S\ref{DLAvoigtfit}). The DLA transmission profile in the right panel is the result of 2-pixel binning along the radial coordinate.
\label{fig:radialprofiles}}
\end{figure*}

To analyze the spatial extent of the DLA, we compare the radial surface brightness profile of the DLA silhouette image, the profile of the background LBG in the continuum image, and the PSF profile in the standard star, Feige 110, image.
These profiles were measured in annular apertures with a spatial interval of 1 pixel ($0.''287$).
The centres of the DLA and LBG profiles were set to be the same and determined by 2D Gaussian fitting to the brightness distribution of the LBG continuum.
For the DLA profile measurement, the LAE near the DLA was masked in the DLA silhouette image although we already excluded most of the wavelengths affected by the LAE (see Table~\ref{tab:wavelengths} for the exact wavelength ranges).
For the LBG profile measurement, a foreground galaxy south of the LBG was masked in the continuum image.
We also applied additional local sky subtraction by measuring the possible sky residual in an annulus with a sufficiently large radius compared to the radii used to measure the profiles. 
The uncertainty was estimated as the standard deviation of the measurements obtained by randomly placing the same annular apertures in the deepest area excluding the DLA/background LBG, the LAE, and the foreground galaxies.
The resulting surface brightness profiles are shown in the left panel of Figure~\ref{fig:radialprofiles}. 
The LBG radial profile perfectly matches the PSF of the KCWI data, and thus the LBG is unresolved.

When we denote the surface brightness profile of the background LBG continuum as $SB_{\rm LBG}$, the profile of the DLA silhouette as $SB_{\rm silhouette}$, and the intrinsic transmission profile of the DLA as $T_{\rm DLA}$, the relationship between them is represented by $SB_{\rm LBG} \times T_{\rm DLA} - SB_{\rm LBG} = SB_{\rm silhouette}$. 
Namely, the intrinsic DLA transmission profile $T_{\rm DLA}$ can be obtained by $T_{\rm DLA} = 1 + SB_{\rm silhouette} / SB_{\rm LBG}$, which is shown in the right panel of Figure~\ref{fig:radialprofiles}.
As found in the right panel of Figure~\ref{fig:radialprofiles}, the DLA transmission in the centre is very consistent with that converted from the covering fraction in the spectral Voigt profile analysis (\S\ref{DLAvoigtfit}), i.e. Transmission = 1 - Covering fraction because the DLA is optically thick.
It might be suggested that the DLA transmission is higher in the outer part.
However, given the fact that the background LBG is unresolved, it would be difficult to discuss the radial profile of the DLA transmission with the current KCWI data.

\section{Discussions}

\subsection{The host galaxy of the DLA}
\label{hostgalaxy}

Although the background LBG has an absolute UV magnitude of $M_{\rm UV}=-21.80$, which is 0.5 mag brighter than a characteristic $M_*$ of LBGs at $z\sim3$ \citep{2022ApJS..259...20H}, it is more than 10 times fainter than typical quasars at $z\sim3$ ($M_*=-24.6$; \citealt{2022ApJS..259...20H}).
The faintness of the background light represents a significant advantage when searching for the host galaxy of a DLA, as it enables closer proximity to the line of sight.
Indeed, we have found an LAE located at $1.''73\pm0.''28$ northeast of the DLA silhouette, corresponding to a projected distance of $12.9\pm2.1$ kpc in the proper coordinate at the LAE redshift $z_{\rm LAE}=3.3433$.
Even for the single absorber case (see Table~\ref{tab:DLAproperties}), the best-fit DLA redshift is $z_{\rm DLA}=3.3419\pm0.0025$ and the redshift difference from the LAE is $\Delta z=0.0014\pm0.0025$, corresponding to the velocity difference of $\Delta v=97\pm173$ km s$^{-1}$, which is consistent with zero, strongly suggesting that the LAE is the host galaxy of the DLA.

The Ly$\alpha$ line flux of the LAE was measured as $(4.78\pm1.27)\times10^{-18}$ erg s$^{-1}$ cm$^{-2}$ in a narrowband image covering the wavelength range [5268.5, 5289.0] (in units of \AA), about twice the line FWHM, and is listed in Table~\ref{tab:laeproperties}, where other properties are also given.
Since the continuum emission of the LAE was not detected significantly, we placed a lower limit on the rest-frame Ly$\alpha$ equivalent width (EW) as $>12$ \AA\ ($3\sigma$) from the limit of the continuum of $<8.9\times10^{-20}$ erg s$^{-1}$ cm$^{-2}$ \AA$^{-1}$ ($3\sigma$) measured in a wavelength range of [5382.0, 5482.0] (in units of \AA).
Given the lower limit of EW(Ly$\alpha$), the galaxy is likely to be classified as an LAE, which are generally defined as galaxies with EW(Ly$\alpha$) $>20$ \AA\ \citep[e.g.,][]{2020ARA&A..58..617O}.

We can constrain the star formation rate (SFR) from the Ly$\alpha$ luminosity.
Assuming the case B recombination, $SFR$(Ly$\alpha$)/(M$_\odot$ yr$^{-1}$) $=L_{\rm Ly\alpha}/(1.1\times10^{42}$ erg s$^{-1}$) \citep[e.g.,][and references therein]{2020ARA&A..58..617O}.
Hence, we obtain $SFR$(Ly$\alpha$) $=0.44\pm0.12$ M$_\odot$ yr$^{-1}$.
Considering a Ly$\alpha$ escape fraction, $f^{\rm esc}_{\rm Ly\alpha}$, which includes the dust attenuation effect, and an IGM transmission, $T^{\rm IGM}_{\rm Ly\alpha}$, the actual SFR should be $SFR=SFR({\rm Ly\alpha})/f_{\rm Ly\alpha}^{\rm esc}/T_{\rm Ly\alpha}^{\rm IGM}$.
Therefore, the obtained $SFR$(Ly$\alpha$) is a strict lower limit because $f^{\rm esc}_{\rm Ly\alpha}\leq1$ and $T^{\rm IGM}_{\rm Ly\alpha}\leq1$.
On the other hand, the $3\sigma$ upper limit of the continuum in the KCWI data reported above corresponds to the upper limit of $SFR$(UV) as $<2.2$ M$_\odot$ yr$^{-1}$ based on the conversion factor of \cite{2012ARA&A..50..531K}.
\cite{2016ApJ...817..161M} reported a more stringent upper limit of $SFR$(UV) as $<0.8$ M$_\odot$ yr$^{-1}$ ($3\sigma$) based on the deeper Suprime-Cam $R_c$-band limiting magnitude.
This limit can be relaxed if there is some dust attenuation, while LAEs are typically dust-poor \citep[e.g.,][]{2020ARA&A..58..617O}.
In summary, the LAE is likely to have an order of $SFR\sim1$ M$_\odot$ yr$^{-1}$.

Assuming that the LAE is in the low-mass regime of a simple extrapolation of the main sequence of SFGs at $z\sim3$ \citep[e.g.][]{2014ApJS..214...15S,2023MNRAS.519.1526P}, the stellar mass is estimated to be $\log_{10}(M_*/{\rm M_\odot})\sim8$.
Based on the SFR-$M_*$ relation of LAEs and low-mass emission line galaxies at $z\sim2$, we obtain the same order of the stellar mass \citep{2016ApJ...817...79H,2024ApJ...964....5C}.
The latest JWST results of low-mass star forming galaxies at $z\sim3$ also support this conclusion quantitatively \citep{2024A&A...683A.184G,2024ApJ...977..133C}.
The dark matter halo mass of an LAE of this order of stellar mass is expected to be of the order of $\log_{10}(M_{\rm halo}/{\rm M_\odot})\sim10$ \citep{2020ARA&A..58..617O}.
The corresponding virial radius at $z=3.34$ is 15 kpc for the density contrast factor of 200.
This is comparable to or slightly larger than the proper distance between the DLA and the LAE of 13 kpc (the impact parameter $b=1.''7$).
Therefore, the DLA can be a dense H~{\sc i} gas in the CGM within the virial radius of the dark matter halo of the LAE.


Another possibility is that the DLA host galaxy is not a single galaxy, but is part of the surrounding galaxy environment.
In fact, several previous studies have found multiple host galaxies for a single DLA, suggesting that DLAs generally belong to the multi-galaxy environment.
\cite{2018MNRAS.479.2126F} detected a CO-rich galaxy at an impact parameter of 117 kpc and a velocity difference $\Delta v = 131$ km s$^{-1}$ from a DLA at $z=2.5832$, and this galaxy forms a galaxy group with another galaxy previously known as the DLA host.
\cite{2019MNRAS.487.5070M} reported a high detection rate ($\approx80$\%) of LAEs within 1000 km s$^{-1}$ from DLAs and with impact parameters between 25 and 280 kpc in the fields of six DLAs at $3.2<z<3.8$.
\cite{2022MNRAS.514.6074N} detected three LAEs and an SFG with weak ISM absorption lines around a DLA at $z=2.431$. 
These four galaxies within 8--28 kpc and $\Delta v\sim40$--340 km s$^{-1}$ form a compact galaxy group hosting the DLA. 
\cite{2023MNRAS.518..305L} report that 30--40\% of H~{\sc i} abosrbers at $z\sim3$--4 are associated with multiple LAEs, including 23 LAEs around nine DLAs ($\log_{10}(N_{\rm HI}/{\rm cm^{-2}})>20.3$), suggesting that high density H~{\sc i} absorbers are located in the outskirt of the CGM of LAEs or in the IGM surrounding these galaxies.

The target DLA of this paper resides in a $\sim30$ comoving Mpc-scale galaxy overdensity \citep{2016ApJ...817..161M}.
Locally, the DLA is surrounded by four galaxies within $30''$ ($=223$ proper kpc at $z=3.34$ in projection; see Figure~\ref{fig:FoVandImages} a), namely two LBGs (LBG1 at a distance of 140 kpc and LBG2 at 209 kpc) and one SMG (71 kpc) and the newly found LAE (13 kpc).
The velocity differences from the DLA at $z=3.3419$ (or $z=3.3433$ same as the LAE; see Table~\ref{tab:DLAproperties}) of these galaxies are $+97$ (0) km s$^{-1}$ for the LAE, $-366$ ($-462$) km s$^{-1}$ for LBG1, $-439$ ($-536$) km s$^{-1}$ for the SMG, and $-2327$ ($-2423$) km s$^{-1}$ for LBG2.
Table~\ref{tab:environment} is a summary of the information.
Three of the four galaxies are within about $500$ km s$^{-1}$ in velocity and 140 kpc in the impact parameter, suggesting that these three galaxies form a galaxy group or a protocluster core.
In this scenario, the DLA can be considered as the H~{\sc i} gas belonging to such a group/protocluster environment.
It will be very interesting to unravel the full structure of this galaxy overdensity hosting the DLA by performing further three-dimensional spectroscopy over a larger area with MUSE and ALMA.

\begin{table}
 \caption{Velocity differences and sky and projection distances of the surrounding galaxies from the DLA. The projection distances are in the physical unit at $z=3.34$.}
 \label{tab:environment}
 \begin{tabular}{lllll}
  \hline
   & & & \multicolumn{2}{c}{Distance} \\
   & Redshift & $\Delta v$ (km s$^{-1}$) & ($''$) & (kpc) \\
  \hline
  DLA & 3.3419 (or 3.3433) & --- & --- & --- \\
  LAE & 3.3433 & $+97$ (0) & $1.''73$ & 13 \\
  LBG1 & 3.3366 & $-366$ ($-462$) & $18.''78$ & 140 \\
  LBG2 & 3.3082 & $-2327$ ($-2423$) & $28.''07$ & 209 \\
  SMG & 3.33554 & $-439$ ($-536$) & $9.''54$ & 71 \\
  \hline
 \end{tabular}
\end{table}

\subsection{The nature of the DLA}
\label{natureofDLA}

Since the background LBG is not resolved in either the current KCWI data or the Suprime-Cam images, we set a lower limit on the emission size.
According to \cite{2016ApJ...817..161M}, the size of an LBG with an absolute magnitude of $M_{\rm UV}=-21.80$ is $1.6^{+2.1}_{-0.9}$ kpc, based on the size-luminosity relation of \cite{2013ApJ...765...68H}.
This size estimate is consistent with the latest observations with JWST \citep{2024MNRAS.533.3724V}.
Taking the lower bound of the size estimate, we obtain a lower bound of the DLA area as $A_{\rm HI}=\pi R_{\rm LBG}^2>1.5$ kpc$^2$.
This limiting area of the DLA is very conservative and two orders of magnitude smaller than the minimum area of $z=2.0$ and 2.5 DLAs estimated by \cite{2022Natur.606...59B}.
Considering the covering fraction, the total H~{\sc i} mass is given by $M_{\rm HI}=m_{\rm p} N_{\rm HI} f_{\rm cov} A_{\rm HI}$, where $m_{\rm p}$ is the proton mass. 
The obtained lower bound of the H~{\sc i} mass of the DLA is $\log_{10}(M_{\rm HI}/{\rm M_\odot})>6.99\pm0.18$ for the single absorber case or $>6.91\pm0.20$ for the absorber at the LAE redshift in the two-component absorber case, respectively.
In any case, a lower bound of the H~{\sc i} mass of the DLA is $\log_{10}(M_{\rm HI}/{\rm M_\odot})\gtrsim7$.
This is again two orders of magnitude smaller than the minimum H~{\sc i} masses of $z=2.0$ and 2.5 DLAs estimated by \cite{2022Natur.606...59B}.
If the DLA of this paper is as massive as those discussed in \cite{2022Natur.606...59B}, namely, a $\sim10^9$ M$_\odot$ H~{\sc i} cloud, it is difficult to consider that the host galaxy of the DLA is the LAE with a dark matter halo of $M_{\rm halo}\sim10^{10}$ M$_\odot$.
In this case, the DLA would belong to the galaxy group/protocluster environment as discussed in \S\ref{hostgalaxy}.
To resolve which DLA host scenario is true, we need to know the actual extent of the H~{\sc i} gas distribution by spatially resolving the background LBG emission and the DLA absorption.

There could be the non-zero residual flux in the bottom of the DLA wavelength range.
We have obtained a $\sim90$\% covering fraction in the Voigt profile analysis in \S\ref{DLAvoigtfit}, suggesting a $\sim10$\% transmission.
The same transmission has also been found in the radial profile analysis in \S\ref{sec:radialprofile}.
Performing $2.''0$ diameter aperture photometry in the narrowband image of the DLA wavelength shown in Figure~\ref{fig:FoVandImages} (d), we have found a $\sim4\sigma$ significance of the residual emission at the position of the DLA/background LBG.
However, there are similarly marginal signals distributed to the west of the DLA/background LBG position and near the edge of one of the FOVs on Day 1 where the noise level is relatively high.
Although the residual emission is not sufficiently robust, if it is real, the transmission of the background LBG continuum through the DLA suggests three possibilities of the internal H~{\sc i} structure: (1) a smooth translucent cloud, (2) a single but offset cloud, and (3) multiple patchy clouds.

In fact, the case (1) a smooth translucent cloud is rejected immediately.
The damping wing profile in the DLA spectrum requires an overall H~{\sc i} column density $N_{\rm HI}\sim10^{21}$ cm$^{-2}$, giving an optical depth of an order of $10^7$, because the H~{\sc i} Ly$\alpha$ cross section is $\sigma_\alpha\sim10^{-14}$ cm$^2$.
Therefore, the cloud must be highly optically thick.
The case (2) an offset cloud predicts a spatial offset of the emission centroids between the LBG continuum at the wavelength not affected by the DLA and the continuum transmitted through the DLA. 
We could not find any significant offset in the current data, although it is inconclusive due to the low signal-to-noise ratio of the transmitted continuum as well as the low spatial resolution. 
We need deeper and higher spatial resolution observations to draw a conclusion. 
One implication of this case for the structure of the H~{\sc i} cloud is that the cloud edge must be sharp enough to reduce the optical depth from $\sim10^7$ to $<1$ within a scale smaller than the cloud. 
The case (3) multiple patchy clouds implies an extremely large variation in H~{\sc i} column density across the gas distribution. 
Although more than an order of magnitude variation in H~{\sc i} column density has been reported by \cite{2022Natur.606...59B}, the variation in column density must be more than seven orders of magnitude for the optical depth to be less than 1 in some parts of the region, while other parts have an optical depth of $\sim10^7$ to make the column density $\sim10^{21}$ cm$^{-2}$.

Either case (2) or (3) requires more than seven orders of magnitude of optical depth contrast in the internal or boundary structure of the DLA.
This optical depth contrast, or equivalently the H~{\sc i} density contrast, can be explained by the situation that the DLA gas, (almost) fully neutral, is surrounded by highly ionized gas.
We call this the ambient ionized medium, which is optically thin for Ly$\alpha$.
Note that the ambient medium is not necessarily less dense than the DLA gas, but only their H~{\sc i} density is more than $10^{7}$ times lower than that in the DLA due to an extremely low neutral fraction.
Given the measured H~{\sc i} column density of $\sim10^{21}$ cm$^{-2}$, the H~{\sc i} volume density in the DLA is $n_{\rm HI,DLA}\sim0.3/l_{\rm DLA,kpc}$ cm$^{-3}$, where $l_{\rm DLA,kpc}$ is the physical scale of the DLA in the unit of kpc. 
Since the H~{\sc i} density contrast is $>10^7$, the H~{\sc i} density in the ambient medium should be $n_{\rm HI,amb}<3\times10^{-8}/l_{\rm DLA,kpc}$ cm$^{-3}$.
Adopting the H~{\sc i} density in the ambient medium under the photoionisation equilibrium with the ionising background radiation (Appendix \ref{ambientmedium}), we find that the condition to reach the H~{\sc i} density contrast of $>10^7$ is $\delta_{\rm amb}+1<23\sqrt{\Gamma_{-12}/l_{\rm DLA,kpc}}$, where $\delta_{\rm amb}$ is the overdensity factor of the ambient medium compared to the cosmic mean density at $z=3.34$ and $\Gamma_{-12}$ is the photoionisation rate normalized by $10^{-12}$ s$^{-1}$, a typical value under the ionising background radiation at $z\sim3$ \citep[e.g.,][]{2013MNRAS.436.1023B}.

As discussed in \S\ref{hostgalaxy}, the DLA can be considered to be a cloud in the CGM located around the virial radius of a dark matter halo of $\sim10^{10}$ M$_\odot$.
The overdensity around the virial radius of a halo should be smaller than its mean value within the radius, which is typically $\sim200$.
In the NFW dark matter profile \citep{1997ApJ...490..493N}, the concentration parameter, $c$, determines the density around the virial radius.
For example, a halo of $M_{\rm halo}\sim10^{10}$ M$_\odot$ at $z\sim3$ expects $c\sim4$ \citep{2014MNRAS.441.3359D} and the dark matter density contrast at the virial radius compared to the cosmic critical density is about 50.
Assuming a constant baryon to matter density ratio, this is an expected baryon overdensity at the virial radius.
Compared to the above condition for ambient overdensity, $\delta_{\rm amb}$, we find $\Gamma_{-12}/l_{\rm DLA,kpc}\gtrsim4$.
Therefore, the extremely high contrast of $>10^7$ in the optical depth in the edge or internal structure of the DLA is realised in the situation that the DLA is surrounded by the highly ionised ambient medium in the CGM, if the DLA scale and the ionising background radiation fulfil this condition.
For example, if the background radiation has a typical intensity of $\Gamma_{-12}\sim1$, then the DLA clouds should be sub-kpc scale, suggesting that a patchy structure composed of such small clouds is more likely.
Given the DLA resides in a galaxy overdensity, the ionising background can be stronger than a typical intensity.
When $\Gamma_{-12}\sim4$, the physical scale of the DLA gas can be $\sim1$ kpc, which is comparable to the typical LBG size.

\subsection{The origin of the O~{\sc i} absorber}

The velocity differences of the DLA and four galaxies in the field compared to the redshift of the O~{\sc i} $\lambda1302$ absorber are listed in Table~\ref{tab:oivelocity}.
The velocity difference between the O~{\sc i} absorber and the DLA is $905$ km s$^{-1}$ (or 1001 km s$^{-1}$ if the DLA has the same redshift as the LAE), which seems large enough to consider that they are different gas components.
On the other hand, the O~{\sc i} absorber is still located within about 1,000 km s$^{-1}$ relative to the galaxies in the field, except for LBG2, suggesting that it is a gas cloud associated with the galaxy group/protocluster consisting of the LAE, LBG1 and the SMG.

The oxygen abundance may give a clue to the origin of the O~{\sc i} absorber.
Assuming the gas is dominated by the neutral phase, the difference between the O~{\sc i} and H~{\sc i} column densities can be converted to the oxygen abundance.
The obtained lower limit is 12$+$log(O/H) $>7.3$.
By comparison with the oxygen abundance of the solar photosphere, 12$+$log(O/H) $=8.69$ \citep{2009ARA&A..47..481A}, we find that the metallicity of the O~{\sc i} absorber is at least 4\% of the solar abundance.
Therefore, the O~{\sc i} absorber is not very primordial, but rather, metal-enriched to some extent, suggesting a possible scenario of an outflow from a galaxy in the environment as the origin of the absorber.

The most massive and active star-forming galaxy in the field is the SMG, which has a molecular gas mass of $3\times10^{11}$ M$_\odot$ and $SFR\sim1,000$ M$_\odot$ yr$^{-1}$ (Inoue et al.\ in preparation).
Let us discuss the possibility that the O~{\sc i} absorber is part of a spherical outflow gas shell from the SMG.
The projected distance at $z=3.34$ between the line of sight (i.e. the location of the O~{\sc i} gas) and the SMG is 71 kpc.
We therefore assume a physical scale of the shell radius ($R_{\rm out}$) of the order of 100 kpc.
The outflow gas mass can be expressed as $M_{\rm out}=4\pi R_{\rm out}^2 \Delta R_{\rm out} \rho_{\rm out} C_{\rm out}$, where $\Delta R_{\rm out}$ is the thickness of the shell, $\rho_{\rm out}$ is the volume mass density of the outflow gas, and $C_{\rm out}$ is the covering fraction of the outflow gas in the shell.
The outflow gas column density is given by $\Delta R_{\rm out} \rho_{\rm out}=\mu m_{\rm p} N_{\rm H}$, where $\mu=1.4$ is the atomic mass weight, $m_{\rm p}$ is the proton mass, and $N_{\rm H}$ is the hydrogen number column density.
Assuming the outflow gas traced by O~{\sc i} to be dominated by the neutral phase, $N_{\rm H}\sim N_{\rm HI} \sim10^{20}$ cm$^{-2}$, which is obtained by the two-component Voigt profile fitting (Table~\ref{tab:DLAproperties}).
Finally, we obtain $M_{\rm out}\sim 1.4\times10^{11} C_{\rm out} (R_{\rm out}/100\,{\rm kpc})^2$ M$_\odot$, which is smaller than the molecular gas mass of the SMG because $C_{\rm out}\leq1$ by definition.

If the angle between the line of sight and the outflow is $\theta$, then the outflow velocity is given by $V_{\rm out}=V_{\rm LOS}/\cos\theta$, where $V_{\rm LOS}$ is the velocity along the line of sight.
For the case of $R_{\rm out}=100$ kpc and the impact parameter of 71 kpc, the angle $\theta\simeq45^\circ$ and $V_{\rm out}\simeq660$ km s$^{-1}$ when $V_{\rm LOS}=466$ km s$^{-1}$.
Hence we obtain the outflow time-scale $T_{\rm out}\sim R_{\rm out}/V_{\rm out}\sim 1.5\times 10^8$ yr and the outflow rate ${\dot M}_{\rm out}\sim M_{\rm out}/T_{\rm out}\sim M_{\rm out} V_{\rm out} / R_{\rm out} \sim 9.3\times 10^2 C_{\rm out}$ M$_\odot$ yr$^{-1}$.
Finally, we find the mass loading factor $\eta={\dot M}_{\rm out}/SFR \sim C_{\rm out}$ for $SFR\sim10^3$ M$_\odot$ yr$^{-1}$.
Although the covering factor is unknown, it should not be too small, given that we have detected the O~{\sc i} absorber by chance.
For example, mass loading factors of a few tens percent are reported for dusty star-forming galaxies (DSFGs) at $z\sim4.5$ and local ultra-luminous infrared galaxies (ULIRGs) with $SFR\sim10^3$ M$_\odot$ yr$^{-1}$ \citep{2020ApJ...905...86S}.
Therefore, $C_{\rm out}\sim0.1$ may be reasonable.
In this case, the outflow gas mass is an order of $10^{10}$ M$_\odot$, which is an order of magnitude larger than those observed in the DSFGs and ULIRGs \citep{2020ApJ...905...86S}.
This contradiction may be caused by the large size of $R_{\rm out}\sim100$ kpc assumed here.
Nevertheless, the outflow gas scenario for the origin of the O~{\sc i} absorber does not appear to be implausible in general.

The LAE is also the possible launcher of the O~{\sc i} outflow because it has the smallest impact parameter of 13 kpc among the galaxies in the field.
If the outflow shell radius is assumed to be $R_{\rm out}\sim30$ kpc, the angle between the line of sight and the outflow is small and $V_{\rm out}\sim 1.1 V_{\rm LOS}\sim 1100$ km s$^{-1}$.
The outflow velocity is very large, close to the upper limit of the distribution of SFGs \citep{2017ApJ...850...51S,2019ApJ...886...29S}.
For this shell radius and an SFR of $\sim1$ M$_\odot$ yr$^{-1}$ (see \S\ref{hostgalaxy}), the outflow mass, timescale, rate, and mass loading factor are $M_{\rm out}\sim1.3\times10^{10}C_{\rm out}$ M$_\odot$, $T_{\rm out}\sim2.7\times10^7$ yr, $\dot{M}_{\rm out}\sim4.7\times10^2C_{\rm out}$ M$_\odot$ yr$^{-1}$, and $\eta\sim4.7\times10^2C_{\rm out}$, respectively.
Since the mass loading factor is known to be anti-correlated with the halo mass \citep[e.g.,][]{2020A&ARv..28....2V,2023MNRAS.525.5868H}, $\eta\sim50$ when $C_{\rm out}\sim0.1$ is a reasonable value for $M_{\rm halo}\sim10^{10}$ M$_\odot$ \citep{2023MNRAS.525.5868H}.
Therefore, either the SMG or the LAE can be the source of the O~{\sc i} outflow.

In addition, there are some hints of other metal absorption features in the wide coverage of the DEIMOS spectrum (see Figure~\ref{fig:widerangespec}).
We will present a full analysis of these possible features, as well as the physical and chemical properties in this galaxy group/protocluster environment in the future.

\begin{table}
 \caption{Velocity differences from the O~{\sc i} $\lambda1302$ absorber.}
 \label{tab:oivelocity}
 \begin{tabular}{lll}
  \hline
   & Redshift & $\Delta v$ (km s$^{-1}$) \\
  \hline
  O~{\sc i} & 3.3288 & --- \\
  DLA & 3.3419 & $+905$ \\
  LAE & 3.3433 & $+1001$ \\
  LBG1 & 3.3366 & $+539$ \\
  LBG2 & 3.3082 & $-1433$ \\
  SMG & 3.33554 & $+466$ \\
  \hline
 \end{tabular}
\end{table}

\subsection{Future prospects}

To study the properties of DLAs and their host galaxies in more detail, statistical research with a large sample is needed.
In particular, to reveal the spatial extent and internal structure of DLAs, it is desirable to have a sample of DLAs found in galaxies as background light sources.
However, such DLAs are very rare, with only four cases (five DLAs) reported to date \citep{2015ApJ...812L..27C,2016ApJ...817..161M,2021ApJ...907..103D,2022Natur.606...59B}.
\cite{2016ApJ...817..161M}, who discovered the target DLA of this paper, discussed the probability of finding DLAs in LBG spectra by assuming a distribution function of H~{\sc i} absorbers in the IGM \citep{2014MNRAS.442.1805I}.
According to their estimate, the probability of finding DLAs in galaxy spectra is an order of 1\%.
Therefore, if we want to obtain $\sim100$ samples, we should observe about $\sim10,000$ LBGs.
If we have a high enough sensitivity to detect lower H~{\sc i} column density systems, such as sub-DLAs and Lyman limit systems, in galaxy spectra, the probability of detection increases significantly thanks to a steep H~{\sc i} column density distribution function \citep[e.g.,][]{2012A&A...547L...1N}.
A large dataset of LBG spectra will be obtained by the Subaru Strategic Program with the Prime Focus Spectrograph (PFS; \citealt{2014PASJ...66R...1T}).
Therefore, we have a good chance of constructing a statistical sample of DLAs in galaxy spectra.
These DLAs found in the galaxy lines of sight will be followed up with Keck/KCWI, VLT/MUSE, and eventually, 30-40 m class extremely large telescopes (ELTs).
Spatial resolution is essential to resolve the DLA structure and therefore the adaptive optics system in optical wavelengths is required.
MAVIS \citep{2020SPIE11447E..1RR} on the VLT will be an excellent instrument to pursue this type of science.
Optical integral field spectroscopy with the Habitable Worlds Observatory\footnote{https://habitableworldsobservatory.org/home} will be the ultimate tool for this science in the future.

\section*{Acknowledgements}

We thank Seiji Fujimoto for discussions in the early stage of this work.
We were supported by JSPS KAKENHI Grant Numbers 21H04489, 22H04939, 23H00131, 24H00002, 24K17095, 25K01038, and 25K01039.
Some of the data presented herein were obtained at Keck Observatory, which is a private 501(c)3 non-profit organization operated as a scientific partnership among the California Institute of Technology, the University of California, and the National Aeronautics and Space Administration. The Observatory was made possible by the generous financial support of the W.\ M.\ Keck Foundation.
The authors wish to recognize and acknowledge the very significant cultural role and reverence that the summit of Maunakea has always had within the Native Hawaiian community. We are most fortunate to have the opportunity to conduct observations from this mountain.
This research made use of Montage. It is funded by the National Science Foundation under Grant Number ACI-1440620, and was previously funded by the National Aeronautics and Space Administration's Earth Science Technology Office, Computation Technologies Project, under Cooperative Agreement Number NCC5-626 between NASA and the California Institute of Technology.

\section*{Data Availability}

The raw data of Keck/KCWI and DEIMOS observations presented in this paper are available on the Keck Observatory Archive\footnote{https://koa.ipac.caltech.edu/cgi-bin/KOA/nph-KOAlogin}.
The raw data of VLT/VIMOS observations presented in this paper are also available on the ESO Archive\footnote{https://archive.eso.org/cms.html}.



\bibliographystyle{mnras}
\bibliography{references} 




\appendix

\section{Sky brightness gradient correction}

\begin{figure}
\centering
\includegraphics[angle=90,width=7cm]{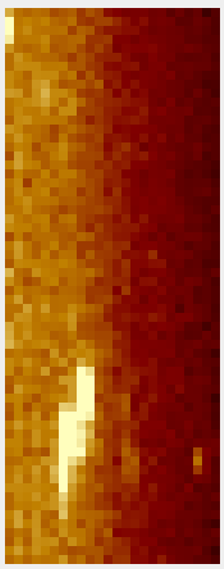}
\includegraphics[angle=90,width=7cm]{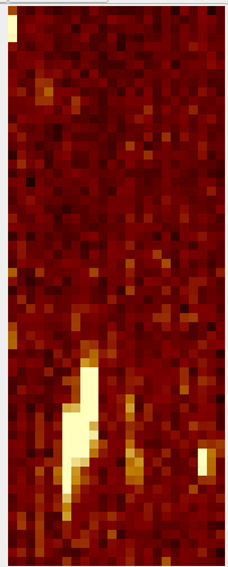}
\caption{Wavelength-integrated images of an example data cube (top) before and (bottom) after flattening the sky background using the method introduced by \protect\cite{2022MNRAS.514.6074N}.
\label{fig:exampleshot}}
\end{figure}

As reported by \cite{2022MNRAS.514.6074N}, the KCWI data cube reduced by the official pipeline has sky brightness gradient as shown in the top panel of Figure~\ref{fig:exampleshot}.
This also depends on the wavelength.
We corrected this wavelength-dependent sky brightness gradient by using the method presented by \cite{2022MNRAS.514.6074N}.
As shown in the bottom panel of Figure~\ref{fig:exampleshot}, the method worked very well to remove the sky gradient.

\section{Two additional KCWI spectral peaks}

\begin{figure}
\centering
\includegraphics[width=7cm]{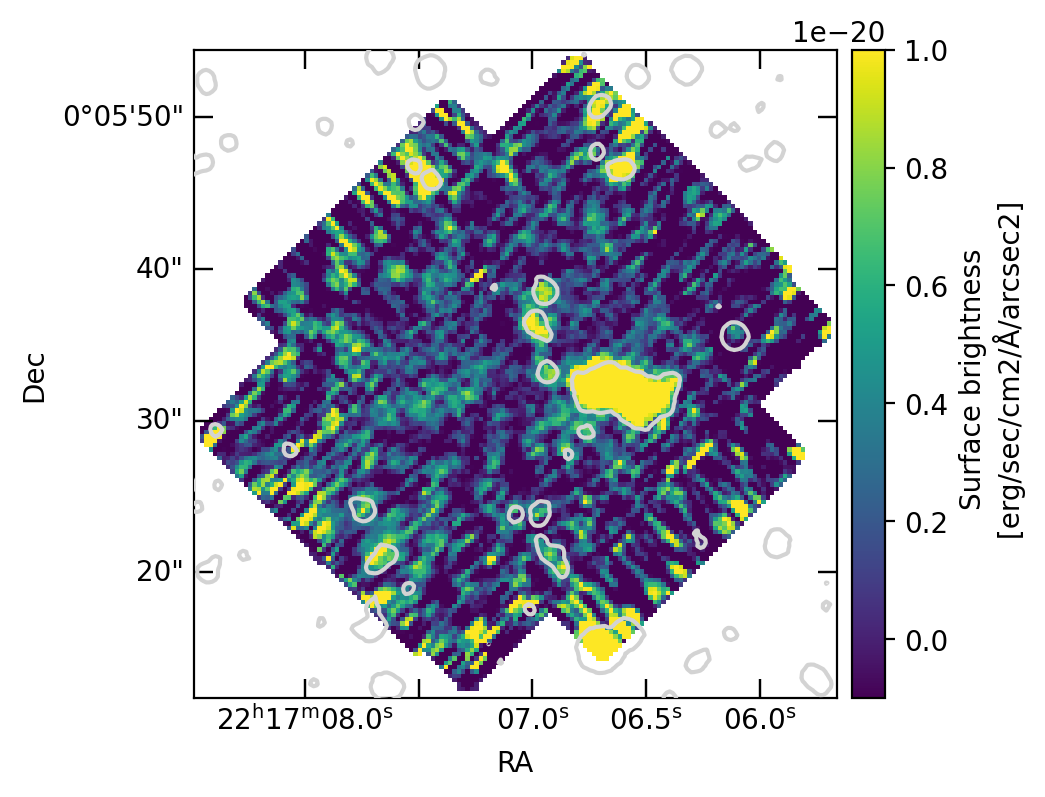}
\caption{The narrowband image integrated over [5210.0 \AA, 5220.0 \AA]. The $1\sigma$ surface brightness level is $3.0\times10^{-21}$ erg s$^{-1}$ cm$^{-2}$ \AA$^{-1}$ arcsec$^{-2}$. The contours show the $5\sigma$ surface brightness level in the Suprime-Cam $i$-band image.}\label{fig:5215Apeak}
\end{figure}

\begin{figure}
\centering
\includegraphics[width=7cm]{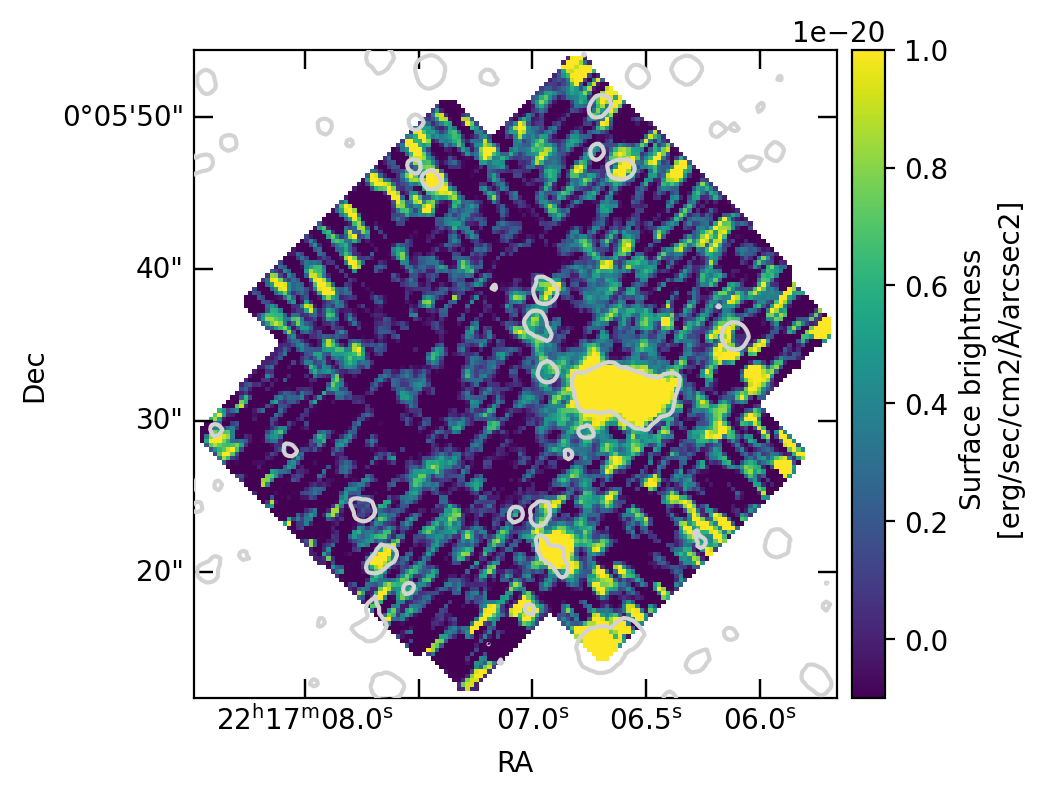}
\caption{Same as Figure~\ref{fig:5215Apeak} but for [5240.0 \AA, 5250.0 \AA]. The $1\sigma$ surface brightness level is $3.3\times10^{-21}$ erg s$^{-1}$ cm$^{-2}$ \AA$^{-1}$ arcsec$^{-2}$.}\label{fig:5245Apeak}
\end{figure}

\noindent
In addition to the peak at 5280~\AA, caused by contamination from the nearby LAE, there are two other peaks at 5215~\AA\ and 5245~\AA\ in the KCWI spectrum shown in the bottom panel of Figure~\ref{fig:DLArangespec}.
We show narrowband images of these two peaks in Figures~\ref{fig:5215Apeak} and \ref{fig:5245Apeak}.
While the weak signals are found at the position of the background LBG, which are likely to cause the spectral peaks, there is no significant detection like the LAE found in the 5280~\AA\ image shown in Figure~\ref{fig:FoVandImages} (c).
Therefore, we conclude that the 5215~\AA\ and 5245~\AA\ peaks are transmission peaks through the foreground H~{\sc i} gas.

\bigskip

\section{The H~{\sc i} density in the ambient medium under photoionisation equilibrium}
\label{ambientmedium}

We consider the situation where the DLA, the (almost) fully neutral gas, is surrounded by the ambient medium, which is highly ionized and optically thin for the Ly$\alpha$ line.
In the ambient medium with the H~{\sc i} density of $n_{\rm HI,amb}$, we assume the photoionisation equilibrium, $\Gamma_{\rm HI} n_{\rm HI,amb} = n_{\rm p} n_{\rm e} \alpha_{\rm B}$, where $\Gamma_{\rm HI}$ is the photoionisation rate, which is an order of $10^{-12}$ s$^{-1}$ under the ionising background at $z\sim3$ \citep[e.g.,][]{2013MNRAS.436.1023B}, $n_{\rm p}$ and $n_{\rm e}$ are the volume number densities of proton and electron in the ambient medium, respectively, and $\alpha_{\rm B}=2.59\times10^{-13}$ cm$^3$ s$^{-1}$ is the H~{\sc i} recombination rate in the so-called case B \citep{2006agna.book.....O}. 
By using the neutral fraction of hydrogen, $x_{\rm HI,amb}=n_{\rm HI,amb}/n_{\rm H,amb}$, where $n_{\rm H,amb}$ is the total hydrogen number density in the ambient medium, $n_{\rm p}=n_{\rm e}=(1-x_{\rm HI,amb})n_{\rm H,amb}\approx n_{\rm H,amb}$ in a highly ionized phase, $x_{\rm HI,amb}\ll1$, and when neglecting the helium contribution. 
Then, we obtain $x_{\rm HI,amb}\approx n_{\rm H,amb} \alpha_{\rm B}/\Gamma_{\rm HI}$.
The mean hydrogen number density at the redshift $z$ in the Universe is given by $n_{\rm H}(z) = \langle n_{\rm H}\rangle_0 (1+z)^3$, where $\langle n_{\rm H}\rangle_0=1.81\times10^{-7}$ cm$^{-3}$ is the mean density in the present Universe ($z=0$) under the assumed cosmological parameters and the mean atomic weight $\mu=1.4$.
The mean density at $z=3.34$ is $1.47\times10^{-5}$ cm$^{-3}$.
When we consider the overdensity factor in the ambient medium, $\delta_{\rm amb}$, compared to the cosmic mean density at $z=3.34$, namely $\delta_{\rm amb}+1=n_{\rm H,amb}/\langle n_{\rm H}\rangle$, the ambient H~{\sc i} density becomes $n_{\rm HI,amb}=x_{\rm HI,amb}\langle n_{\rm H} \rangle (\delta_{\rm amb}+1)=(\delta_{\rm amb}+1)^2 \langle n_{\rm H} \rangle^2 \alpha_{\rm B}/\Gamma_{\rm HI}=5.60\times10^{-11} (\delta_{\rm amb}+1)^2/\Gamma_{-12}$ cm$^{-3}$, where $\Gamma_{-12}$ is $\Gamma_{\rm HI}$ normalized by $10^{-12}$ s$^{-1}$.


\bsp	
\label{lastpage}
\end{document}